\documentclass[11pt]{article}

\usepackage{amsmath, amsfonts}
\usepackage{graphicx}
\usepackage{ulem}
\usepackage{color}
\usepackage{url}
\usepackage{soul}
\usepackage[makeroom]{cancel}

\newif\ifblog
\newif\iftex
\blogfalse
\textrue

\def\emph#1{\textit{#1}}

\makeatletter
\renewcommand\@biblabel[1]{#1.}
\makeatother

\def\P{{\mathbb P}}
\def\E{{\mathbb E}}

\def\N{{\mathbb N}}

\def\R{{\mathbb R}}
\def\C{{\mathbb C}}
\def\D{{\mathbb D}}

\newtheorem{theorem}{Theorem}
\newtheorem{lemma}[theorem]{Lemma}

\newtheorem{proposition}[theorem]{Proposition}
\newtheorem{example}[theorem]{Example}
\newtheorem{remark}[theorem]{Remark}
\newtheorem{assumption}[theorem]{Assumption}

\newcommand\numberthis{\addtocounter{equation}{1}\tag{\theequation}}

\title{Weak Convergence of Path-Dependent SDEs in Basket CDS Pricing with Contagion Risk
}

\author{Yao Tung Huang,
\thanks{Magnum Research Limited, Entrepreneurship Center, Hong Kong University of Science and Technology, Clear Water Bay, Kowloon, Hong Kong. don.huang@magnumwm.com.}
\and {Qingshuo Song,
\thanks{Department of Mathematics, City
University of Hong Kong, 83 Tat Chee Avenue, Kowloon Tong, Hong
Kong. song.qingshuo@cityu.edu.hk.}
}
\and {Harry Zheng
\thanks{Department of Mathematics,
Imperial College, London SW7 2AZ, UK. h.zheng@imperial.ac.uk}}
}

\date{}

\begin{document}

\maketitle

\begin{abstract}
We investigate the computational aspects of the basket CDS pricing
with counterparty risk
under  a credit contagion model of multinames.
This model enables us to capture the systematic volatility increases
in the market triggered by a particular bankruptcy.
The drawback of this problem is its analytical complication
due to its path-dependent functional, which bears
a potential failure in its convergence of numerical approximation
under standing assumptions.
In this paper we find sufficient conditions
for the desired convergence of the functionals
associated with a class of path-dependent stochastic differential equations.
The main ingredient is to identify the weak convergence of the 
approximated solution to the underlying path-dependent stochastic differential equation.
\end{abstract}

\smallskip
\noindent \textbf{Keywords.} Path-dependent SDE, weak convergence, correlated  first-passage times, basket CDS, contagion risk,  counterparty risk.

\bigskip
\bigskip
\noindent \textbf{JEL Classification.} C02, G12.
	
\noindent \textbf{Mathematics Subject Classification (2010).} 60F05, 60H30, 60J60, 91G40.
\section{Introduction}
%
It is well known that the first-passage time of a drifting Brownian
motion crossing a deterministic level is an inverse Gaussian random
variable, as its running maximal  process
of a drifting Brownian motion can be characterized with the reflection
principle and Girsanov theorem. This result has been applied,
among others, in studying the default time of a firm with a structural
framework in which a firm defaults at the first time when its asset
value falls below its liability value. The structural model is acclaimed to
have  strong economic foundation and is one of the most popular models
used in pricing single-name credit derivatives.

It is natural to ask whether one can also characterize the joint distribution
of  first-passage times of correlated drifting Brownian motions
crossing some deterministic levels. When there are two correlated
Brownian motions, \cite{Iyengar1985,Metzler2010} find
the joint distribution of  first-passage times, which is an infinite
sum of modified Bessel functions of the first kind. Little is
known for  first-passage times of three or more correlated Brownian
motions. This limitation  makes it difficult to study the default
times of multiple names with a first-passage time structural model
and is one of the reasons that intensity-based reduced-form models
are  used in pricing portfolio credit derivatives.
After all, intensity models give
more analytic tractability but do not provide economic reasons why
firms default (see \cite{J.W.Gu2013, Zheng2013} for
some recent results in this direction and the references therein).

In this paper we discuss the pricing of a basket CDS with counterparty (CDS writer) risk. The default times of names in a reference portfolio and that of the
counterparty are modeled by a first-passage time structural model
and correlated Brownian motions. Furthermore, we include contagion risk in the model, in which the default of a name in the reference portfolio causes a jump increase of  volatility of the counterparty, which increases the default probability of the counterparty. 
The aforementioned  model, for instance   \cite{Haworth2007}, 
incorporates  both counterparty risk and contagion risk, whereby it
can realistically explain the severe difficulties experienced by some seemingly default-remote banks underwriting super senior tranche CDOs (collateralized debt obligations) during the financial crisis of 2007-08.  

Since the joint distribution of correlated default times in a structural model
is unknown, we
compute the price of a basket CDS  by the Euler scheme for SDEs and
the Monte Carlo simulation.
As an effective computational tool,
Euler scheme has been widely adopted in the credit risk computing
for its simplicity and robustness.
Is it possible that a price computed from these
``robust'' algorithms could be actually over-valued in the trillions dollar market? 

 As we will illustrate in Example~\ref{e:02}, the answer is affirmative in general.
Then, is there a set of broad  conditions, 
which ensures the computation going to the correct value?
In this paper, we aim to reveal the reasons for the
possible mispricing, and further to provide a rigorous justification
on the mispricing scenarios. To illustrate the idea, we take a simplified example which indeed motivates our general setting in Section~\ref{sec:setting}.

\begin{example}
 \label{e:01}
 Let  $V_{0}$ be 
 the firm value process of 
 the CDS writer, and
 $\{V_{i}: i=1, \ldots, k\}$ those of 
 $k$ basket reference
names.
We assume that the default time of the firm $i$  is given by
$\tau_{i} =\inf\left\{ t>0:V_{i}\left(t\right)\leq 1\right\}$, 
$i=0,1,\ldots,k$
Moreover,
we assume
the volatility of firm $i$
follows
$\bar \sigma_{i} \cdot (1 + \alpha(t))$, where
 \begin{equation} \label{eq:dXt}
\begin{array}{ll}
\alpha(t) = \sum_{j=1}^{k} I(\tau_{j} \le t)
\end{array}
\end{equation}
is the total 
default number of reference names 
by time $t$.
The main interest of this this paper is to compute a 
premium rate $\hat{c}$ per annum in the form of
\begin{equation}
\label{eq:hc01}
\hat c = \frac{\mathbb E [f_{1}(\tau_{0},  \ldots, \tau_{k})]}
{\mathbb E[ f_{2}(\tau_{0},  \ldots, \tau_{k})]}
\end{equation}
for some discontinuous functions $f_{1}$ and $f_{2}$ with domain $\mathbb R^{k+1}$, see also the exact formula \eqref{eq:chat_comp}.

Given that Euler scheme with step size $h$ in the approximation of 
the underlying firm values and further their default times ,
denoted by $V_{i}^{h}$ and $\tau_{i}^{h}$ respectively, 
one can approximate premium rate $\hat c$ by replacing 
$\tau_{i}$ by $\tau_{i}^{h}$ in the formula \eqref{eq:hc01}, 
denoted by $\hat c^{h}$.
Our question is that whether the convergence $\hat c^{h} \to \hat c$ holds
as $h\to 0$?
\end{example}

Regarding the numerical schemes of SDEs, there have been extensive research on the topics of both weak and strong convergences, see  
\cite{Higham2002,Hutzenthaler2012,P.E.Kloeden2011,Mao2007,Yan2002}
 and the references therein for excellent expositions. In particular, the book \cite{P.E.Kloeden2011}  introduces a systematic and rigorous treatment
  to the numerical approximation of the various types of SDEs. To the best of our knowledge, essentially all the above literatures on numerical SDEs study the convergence at fixed grid-points in Markovian settings. Back to Example~\ref{e:01}, the volatility is path-dependent and the presence of first-passage times requires information on the whole path.

In this regard, we turn our attention to study associated
mappings from Skorohod path space $\mathbb D$ to real numbers.
For instance, the first passage time $\tau_{i}$ can be rewritten by
$\tau_{i} = \pi(V_{i})$ where $\pi: \mathbb D \to \mathbb R$ is a mapping
defined by $\pi(x) = \inf\{t>0: x(t) \le 1\}$. Similar idea can be applied
to rewrite the premium rate by
$\hat c = \mathbb E[ F_{1}(V) ] / \mathbb E[ F_{2}(V) ]$, where
each $F_{i}$ is a functional on $\mathbb R^{k+1}$-valued path space
$\mathbb D^{k+1}$ with its argument
$V = (V_{i})_{i=0, \ldots, k} \in \mathbb D^{k+1}$. To this end, our work
can be clearly divided into two steps:
The convergence $\hat c^{h} \to \hat c$
shall be true by the continuous mapping theorem, if
\begin{enumerate}
 \item[(Q1)] $V^{h}$ converges to $V$ in distribution 
  (see Section~\ref{sec:conv}) 
  , denoted by $V^{h} \Rightarrow V$;
 \item[(Q2)] $F_{1}$ and $F_{2}$ are continuous almost surely at $V$
 with respect to Skorohod topology.
\end{enumerate}

Regarding (Q1),
the weak limit theorems for the whole path in non-Markovian setting can
be found in \cite{Kurtz1991}, and
our approach for the weak convergence $V^{h} \Rightarrow V$
is also closely related to \cite{Kurtz1991}. However, \cite{Kurtz1991}
establishes the convergence based on the continuity assumption
of coefficient functions, while the volatility of Example~\ref{e:01}
is not continuous as a mapping on a path space, and hence their
result can not be directly applied here. The main reason for the
discontinuity is due to the dependence on the number of defaults $\alpha$,
see also the tangency problem of \cite{Kushner2001}. Nonetheless, $\alpha$ as a function on a path space is 
almost surely continuous with respect to the probability induced by $V$,
see Example~\ref{e:03} for  details.
As such, we bravely attempt to show the
desired weak convergence under almost sure continuity assumption
of coefficient functions, i.e.
\begin{enumerate}
 \item [(H)] As a mapping from path space $\mathbb D^{k+1}$ to $\mathbb R$, the function $\alpha$  is continuous under Skorohod topology almost surely with respect to
$\mathbb P V^{-1}$.
\end{enumerate}
In the above, $\mathbb P V^{-1}$ refers to the probability measure 
on $\mathbb D^{k+1}$ induced by $V$, see more explanation in
Section~\ref{sec:conv}.
 Although (H) is enough for our purpose to cover our motivated example,
it is still inappropriate since the unknown solution $V$ shall not be
included in the assumption. This leads to Assumption~\ref{a:cont_phi},
which serves the same role as (H) with the help of
Assumption~\ref{a:nondegenerate}.

To this end, it is inevitable to go through
the entire procedure and carefully
reexamine all the necessary steps in the weak convergence.
Firstly, we show the tightness of the discrete Euler processes, and deduce the convergence of approximating processes to a limiting process almost surely by the Skorohod representation theorem. Secondly, we claim  the continuity of the limiting process, which plays a crucial role in the proof, see Remark~\ref{Re:Skorohod_top_C}. Finally, we complete the proof by
showing that
the limiting process is the weak solution of the underlying SDE.

Regarding (Q2), provided the completion of (Q1), we shall
show the convergence in distribution
$f_{i}(\tau^{h}_{0},  \ldots, \tau^{h}_{k}) \Rightarrow
f_{i}(\tau_{0},  \ldots, \tau_{k})$ of \eqref{eq:hc01}.
Although the form of $f_{i}$ corresponding to the pricing formula
\eqref{eq:chat_comp} is complicated, it's enough to examine
the weak convergence on the following two simple quantities
by the continuous mapping theorem (CMT):
\begin{enumerate}
 \item (The convergence in single name risk)
 One shall verify $I(\tau^{h}_{i}>t) \Rightarrow I(\tau_{i}>t)$ for arbitrary $i$
 and $t$. Applying CMT, it's sufficient to show
 $\mathbb P(\tau_{i} = t) = 0$ for any $t$, i.e.
 all underlying firms has zero probability to get into default
 at a particular time. This is guaranteed by
 non-degeneracy Assumption~\ref{a:nondegenerate}.

 \item (The convergence in counter-party risk) One shall verify
 $ I(\tau^{h}_{0}>\tau^{h}_{i}) \Rightarrow I(\tau_{0}>\tau_{i})$ for each
 $i\ge 1$. Again by CMT, it's enough to show
 $\mathbb P\{\tau_{0} = \tau_{i}\} = 0$, i.e. there shall be
 zero probability that two companies default simultaneously.  It boils down two sufficient conditions in turn:
 if either (a) CDS writer is independent to all reference names; or
 (b) CDS writer is not perfectly correlated to all reference names and the volatilities are all piecewise constants, then the above convergence
 holds.
\end{enumerate}

To close up the introduction, our contribution is summarized as
follows. We establish the convergence for the approximation
of CDS pricing, see Theorem~\ref{t:main}.
As illustrated above, the major mathematical
difficulty compared to the existing literature stems from the following
added features in our model: (a) the contagion risk due to
the dependence of the volatility
in total number of defaults; (b) single/counter-party default risk
in terms of first-passage times. As of any other weak limit results
on the path space, our result has to be established by many assumptions,
for which we add
careful explanations why they are needed, see
Section~\ref{sec:asm}. As a result, the seemingly cumbersome
assumptions can be reduced to a simple condition in 
Example~\ref{e:01}: If $\bar \sigma_{i}>0$ for all $i$ and $V_{0}$ is not perfectly correlated to any of $\{V_{i}, i= 1, \ldots k\}$, then $\hat c^{h} \to\hat c$ holds.

The paper is organized as follows. Section 2 presents the
problem setup and our main result, which is
to cover rather general scenarios than Example~\ref{e:01}.
Note that some notions used in Introduction are also slightly
extended in an obvious way.
Section 3 includes the technical proof of the main result.
Section 4 includes further discussions related to this work.

\section{Main Results}\label{sec:setting}
\subsection{Problem Setting}
Let $\hat T = T +1$ for some positive constant $T$.
Denote by $\mathbb{D}^{n}$ the space of c\`adl\`ag
functions, i.e. functions that are right continuous with left limits defined 
on $[0,\hat T]$ taking values in $\mathbb{R}^{n}$. $\mathbb{D}^1$ is abbreviated as $\mathbb{D}$.  Let  $\left(\Omega,\mathcal{F},\{\mathcal{F}_t\},
\mathbb{P} \right)$ be a filtered probability space satisfying the usual
conditions, on which a standard $k+1$ dimensional Brownian motion $W=(W_0,W_1,\dots,W_k)^T$
is defined. Here, $x^{T}$ is the transpose of $x$. Suppose that $\mu(\cdot,\cdot):\D^{k+1} \times  [0,T] \mapsto \R^{k+1}$ and $\sigma(\cdot,\cdot):\D^{k+1} \times [0,T] \mapsto \R^{(k+1)\times (k+1)}$ are nonanticipating in the sense that $\mu(x,t) = \mu(x(\cdot\wedge t),t)$ and $\sigma(x,t) = \sigma(x(\cdot\wedge t),t)$ for all $t\geq 0$ and $x\in\D^{k+1}$  (see \cite{Kurtz1991} for details). We consider $k+1$ companies with the firm value processes $V:=(V_0,V_1,\dots,V_k)^T$ satisfying
\begin{equation}
 \label{eq:dVt}
dV\left(t\right)=diag\left(V\left(t\right)\right)\left(\mu\left(V,t\right)\thinspace dt+\sigma\left(V,t\right)\thinspace dW\left(t\right)\right),\quad t\geq 0,
\end{equation}
where the constant $V\left(0\right)$
is the firm value at $t=0$, $diag\left(V\left(t\right)\right)$ is
a $\left(k+1\right)\times\left(k+1\right)$ diagonal matrix with
diagonal elements $V_{i}\left(t\right)$, $i=0,1,\dots,k$ and $\mu =(\mu_0,\mu_1,\dots,\mu_k)^T$
and $\sigma=(\sigma_{ij})_{0\leq i,j\leq k}$  represent asset appreciation and volatility rates, respectively. The product of $\sigma$
and its transpose reflects the covariance between the movements in
the asset values of firms, thus playing a critical role in determining
the dependence structure among the firm values.

Let $\phi$ stand for $\mu$ and $\sigma$.  We impose the following
structure to $\phi = \mu, \sigma$:
For any $x\in \mathbb{D}^{k+1}$, $\phi(x,\cdot)$ can be decomposed into a continuous deterministic process $\phi_c(\cdot)$ and a pure jump process $\phi_J(x,\cdot)$ as follows:
\[
\phi\left(x,t\right) = \phi_{c}\left(t\right)+\phi_{J}\left(x,t\right) = \phi_{c}\left(t\right)+\sum_{i=1}^{\mathcal{N}^{\phi}(x,t)}J_{i}^{\phi}(x),\quad 0\leq t\leq \hat T,
\]
where functions $\mu_J(\hbox{ respectively } \sigma_J):\D^{k+1} \times[0,T]\mapsto \R^{k+1}$ (respectively $\R^{k+1}\times \R^{k+1}$) and $\mathcal{N}^{\phi}:\D^{k+1} \times[0,T] \mapsto \N$ are measurable and nonanticipating in the sense that $\phi_J(x,t)=\phi_J(x(\cdot\wedge t),t)$ and $\mathcal{N}^{\phi}(x,t)=\mathcal{N}^{\phi}(x(\cdot\wedge t),t)$.
The function $J_i^{\mu}(\hbox{ respectively }J_i^{\sigma}):\D^{k+1} \mapsto \R^{k+1}(\hbox{ respectively }\mbox{$\R^{k+1}\times \R^{k+1}$})$ is measurable for $i=1,2,\dots.$ In the above, $\mathcal{N}^\phi(x,t)$ and $J_i^{\phi}(x)$ represent the number of jumps of $\phi$ up to time $t$ and the jump size for the $i$th jump, respectively. Moreover, the processes $t\mapsto \phi(V,t)$  and $t\mapsto \mathcal{N}^{\phi}(V,t)$ are $\{\mathcal{F}_t\}_{t\geq 0}$-adapted.

Without loss of generality, let $V_{0}$ be the firm value process of a CDS writer and
$V_{i}$, $i=1,2,\dots,k$, be the firm value processes of the
companies in the reference portfolio. Each
company has an exponential default barrier $L_{i}$
in the form of
\[
L_{i}\left(t\right)=K_{i}e^{\gamma_{i}t},\quad t\geq 0, i=0,1,\dots,k,
\]
where $\gamma_i$ and $K_{i}$ are nonnegative constants. The default
time for  company $i$ is defined by
\[
\tau_{i}:=\inf\left\{ t>0:V_{i}\left(t\right)\leq L_{i}\left(t\right)\right\}\wedge \hat T, \ i = 0, 1, ..., k
\]
the first time the firm value falls below the default barrier.
We denote by $\{ \tau_{\left(i\right)}:i=1,2,\dots,k\} $ the order statistics of $\left\{ \tau_{i}:i=1,\dots,k\right\}$, i.e. $\tau_{\left(i\right)}$ is the $i$th default
time among $k$ companies in the reference portfolio.

Note that the definition of the default time $\tau_{i}$ is truncated by
$\hat T = T+1$, but not by the maturity $T$ only for its convenience.
The advantage is that,
our work is reduced to C\`adl\`ag space from infinite time interval to
a finite time interval, while we can keep the probability of $\{\tau_{i} = T\}$
as zero under non-degenerate condition. This feature will be used
to show the weak convergence involved with ${\bf 1}(\tau_{(i)} \le T)$, 
while preserve the structure of the swap rate defined in \eqref{eq:chat_comp}.

In pricing derivative securities with the structural model, it is normally assumed that the drift coefficient $\mu$ of $V$ in (\ref{eq:dVt}) is equal to the risk-free interest rate $r$ in a risk-neutral setting. We do not insist that $\mu$ equal to $r$ in this paper as the firm value is not a traded asset and cannot be hedged with the no-arbitrage and martingale representation argument. The firm value process $V$ is a measure that may have close relation with traded assets but is mainly used to define default events. All results still hold if $\mu$ is replaced by $r$. For the sake of simplicity, the risk-free interest rate is  assumed to be a positive constant $r$. The extension to stochastic interest rate model can be done under the framework of this paper (see Remark~\ref{re:stochastic_r} for details).

A basket CDS is an insurance product in which the underlying is a portfolio of defaultable companies and the writer (seller) of the $i$th default CDS promises to pay  $1-\delta_i$ to the buyer of the insurance at the $i$th default time $\tau_{(i)}$  if that happens before the maturity time $T$ of the contract, in return the buyer of the $i$th CDS agrees to pay the writer a premium fee at rate 
$\hat{c}_i$ 
 per annum on each of pre-specified dates $\{ 0<t_{1}<t_{2}<\cdots<t_{m}=T\}$ as long as the $i$th default has not occurred. (In fact, the triggering time for the basket CDS does not have to be the default time of a company, it can be any predefined event). If the writer defaults before the maturity of the contract or the $i$th default time, then the CDS contract   terminates and there are no further cash flows.
The risk neutral swap rate 
$\hat{c}_i$
is given by
\begin{equation}
\label{eq:chat_comp}
\hat{c}_i
=\frac{\mathbb{E}\left[e^{-r\tau_{\left(i\right)}}\left(1-\delta_{i}\right)\mathbf{1}\left(\tau_{\left(i\right)}\leq T\right)\mathbf{1}\left(\tau_{0}>\tau_{\left(i\right)}\wedge T\right)\right]}{\mathbb{E}\left[\sum_{j=1}^{m}e^{-rt_{j}}\Delta t_{j}\mathbf{1}\left(\tau_{\left(i\right)}>t_{j}\right)\mathbf{1}\left(\tau_{0}>t_{j}\right)\right]},
\end{equation}
where $\Delta t_j = t_j - t_{j-1}$, $j=1,\dots,m$, $t_0=0$, and $\mathbf{1}\left(\cdot\right)$
is the indicator  function which equals 1 if an event occurs and 0 otherwise. The values of $\tau_{0}$ and $\tau_{\left(i\right)}$ are dependent
on the realized path of the process $V$. The evaluation of 
$\hat{c}_i$
in \eqref{eq:chat_comp} involves the
expectations of path-dependent functionals, which may naturally be computed with the Monte Carlo and the Euler approximation method.

Let the time interval $[0,T]$ be partitioned into $N$ equally spaced subintervals with grid points $t^h_n=nh$, $n=0,\dots,N$, $h=T/N$, and let $V^{h}$, valued in $\mathbb{R}^{k+1}$, be the Euler approximating
process for $V$, defined recursively by
\begin{equation*}
V_{n+1}^{h}:=V_{n}^{h}+diag\left(V_{n}^{h}\right)\left(\mu_n^h h+\sigma_n^h(W(t_{n+1}^h)-W(t_{n}^h))\right),\quad n=0,\dots,N-1,
\end{equation*}
where $\mu^h_n:=\mu(V^h,nh)$ and $\sigma^h_n:=\sigma(V^h,nh)$.
 In the rest of the paper, for ease of the notation complexity without ambiguity, we retain
the notation $(V^{h}, \mu^{h}, \sigma^{h})$ to denote the piecewise constant interpolation of sequences 
$\{(V_{n}^{h}, \mu_{n}^{h}, \sigma_{n}^{h}): n = 0, 1, \ldots N\}$, i.e. 
\begin{equation}
 \label{eq:Vh01}
V^{h}\left(t\right)=V_{n}^{h},\quad \mu^{h}\left(t\right)=\mu_{n}^{h}, \quad \sigma^{h}\left(t\right)=\sigma_{n}^{h},\quad t\in\left[nh,\left(n+1\right)h\right). 
\end{equation}
Random variables $V_n^h$ are $\mathcal{F}_n^h$-measurable, where $\mathcal{F}_n^h:=\mathcal{F}_{t_n^h}$ is the information available at time $t_n^h$. Since $W$ is a Brownian motion, 
 without changing
their distributions,
we may 
generate $\{ V_{n}^{h}:n=1,2,\dots,N\} $ by the recursive formula
\begin{equation}
\label{eq:V_n+1^h}
V_{n+1}^{h}=V_{n}^{h}+diag\left(V_{n}^{h}\right)\mu_n^h h+diag\left(V_{n}^{h}\right)\sigma_n^h\sqrt{h}Z_{n+1},
\end{equation}
where $Z_{n}$, $n=1,\dots,N$, are independent $k+1$ dimensional
standard normal variables and $Z_l$ are independent of the filtration ${\cal F}^h_{n}$ for $l>n$ and $n=1,\ldots,N-1$.

Corresponding to the Euler approximating process $V^{h}$, the annual swap rate 
$\hat{c}_i^h$
has the same form as that in \eqref{eq:chat_comp} except $\tau_0$, $\tau_i$ and $\tau_{\left(i\right)}$ are replaced by $\tau_0^h$, $\tau_i^h$ and  $\tau_{\left(i\right)}^{h}$, respectively.
In this paper we explore under what conditions
we have
\begin{equation}
\label{Q:c_h_to_c}
\lim_{h\to0}
\hat{c}_i^h
=
\hat{c}_i
.
\end{equation}

\subsection{The Main Results}
In this part, we present the main result after several assumptions.
\begin{assumption}
 \label{a:bs}
 With some positive constant $K$ for all $x\in \mathbb{D}^{k+1}$ and
  $0\leq  t_{1},t_{2}\leq \hat T$, for $\phi = b, \sigma$
\begin{eqnarray*}
\left|\phi_{c}\left(t_{1}\right)-\phi_{c}\left(t_{2}\right)\right| & \leq & K\left|t_{1}-t_{2}\right|^{1/2},\\
\left| \mathcal{N}^{\phi}(x,T)\right| & \leq & K,\\
\left| J^{\phi}_i(x)\right| & \leq & K \text{ for all $i$.}
\end{eqnarray*}
\end{assumption}
\begin{assumption}
 \label{a:nondegenerate}
 $\sigma(x,t)$ satisfies the uniform nondegeneracy condition, i.e. $\sigma(x,t) \sigma(x,t)^T\geq \lambda I$ for all $x\in \mathbb{D}^{k+1}$ and $t\in[0, \hat T]$ for some $\lambda>0$.
\end{assumption}
We next define $\pi:\mathbb{D}\times\mathbb{D}\mapsto\mbox{\ensuremath{\mathbb{R}}}$
as
\begin{equation}
 \label{eq:pi01}
 \pi\left(x,l\right)
:=
\inf\left\{ t>0:x\left(t\right)\leq l\left(t\right)\right\}\wedge \hat T,
\end{equation}
which is the first time of the c\`adl\`ag function $x$ hitting the barrier $l$.
Let $\C^n$ be the collection of continuous functions defined on $[0, \hat T]$ taking values in $\mathbb{R}^{n}$. $\mathbb{C}^1$ is abbreviated as $\mathbb{C}$. For $i=0,1,\dots,k$,  define two disjoint subsets of the space $\C$ as
\begin{align*}
C_{1}^{i} & =\left\{ x\in \C:\pi\left(x,L_{i}\right)<T\thinspace\thinspace\text{and}\thinspace\thinspace\inf\left\{ t>\pi\left(x,L_{i}\right):x\left(t\right)<L_{i}\left(t\right)\right\} =\pi\left(x,L_{i}\right)\right\} ,
\end{align*}
and
\[
C_{2}^{i}=\left\{ x\in\mathbb{C}:\pi\left(x,L_{i}\right)\geq T\right\} .
\]

\begin{assumption}
 \label{a:cont_phi}
The mappings $x\mapsto \mu(x,\cdot)$ and $x\mapsto \sigma(x,\cdot)$
are continuous under Skorohod topology at the set
$\{x\in \C^{k+1}:x_i\in C_1^i \cup C_2^i,i=0,1,\dots,k\}$.
\end{assumption}

\begin{assumption}
 \label{a:zero_corr}
For all $t\geq0$, $\left(\sigma\left(t\right)\right)_{0i}=\left(\sigma\left(t\right)\right)_{i0}=0$, a.s. for $1\leq i\leq k$.
\end{assumption}

\begin{assumption}
 \label{a:ps_const}
$\sigma$ is piecewise constant
almost surely, i.e. for strictly increasing stopping time sequence $\left\{ \theta_{0},\theta_{1},\dots\right\} $
such that $\theta_0 = 0$ and $\lim_{n\to\infty}\theta_n = 
 \hat T
$, the process $\sigma$
is in the following form
\[
\sigma\left(t\right)=\sum_{i=1}^{\infty}\sigma_{i}\mathbf{1}\left(\theta_{i-1}\leq t< \theta_{i}\right),\quad\text{a.s.,}
\]
where $\sigma_{i}$ is a nonsingular $\left(k+1\right)\times\left(k+1\right)$
matrix, measurable with respect to $\mathcal{F}_{\theta_{i-1}}$, $i=1,2,\dots$
\end{assumption}
We now state the main result of the paper, and the explanations are
immediately followed by Section~\ref{sec:asm}. A short remark on
convergence in distribution is also included in Section~\ref{sec:conv}.
\begin{theorem}
\label{t:main}
Let $V$ be the $k+1$ dimensional $\mathcal{F}_{t}$-adapted
continuous process of the form \eqref{eq:dVt}  and $V^{h}$ be 
Euler approximating process  for $V$ given by \eqref{eq:Vh01}. 
 If Assumptions \ref{a:bs}, \ref{a:nondegenerate}, and \ref{a:cont_phi} hold, then $V^{h}$ converges to $V$ in distribution.
If, in addition, either Assumption \ref{a:zero_corr} or \ref{a:ps_const} holds, then ${\displaystyle \lim_{h\to0}}
\hat{c}_i^h
= 
\hat{c}_i$ 
for all $i=1, 2, \ldots, k$.
\end{theorem}

\subsection{Discussions on Assumptions with
its Application to Example~\ref{e:01}} \label{sec:asm}
In this part, we discuss the above assumptions combined with Example~\ref{e:01} to illustrate the main result.

\subsubsection{Discussion on Assumption~\ref{a:bs}}
Assumption~\ref{a:bs} is imposed to guarantee the existence and uniqueness of a solution $V$ to \eqref{eq:dVt}.  As one shall further
note from Assumption~\ref{a:bs}, $\phi(x,\cdot),\phi_c(\cdot)$ and $\phi_J(x,\cdot)$ are all bounded by some constant $K$. (In this paper, $K$ is a generic constant
whose value may change at each line).  We may relax the Holder-$1/2$ continuity of $\phi_c$ by a condition $|\phi_c(t_1)-\phi_c(t_2)|\leq g(|t_1-t_2|)$, where $g$ is a bounded function satisfying $g(0)=0$ and $g(s)$ tends to 0 as $s$ tends to 0.

\subsubsection{Skorohod space}
To discuss other assumptions, we shall first state the Skorohod metric for $\mathbb{D}^{k+1}$ space and related notions, which are adopted by
\cite{Billingsley2009}.
Define a uniform metric on $\mathbb{D}^{k+1}$ by
\begin{equation}
 \label{eq:ufm}
\left\Vert x-y\right\Vert =\sup_{t\in\left[0,T\right]}\left|x\left(t\right)-y\left(t\right)\right|,\forall x,y\in \mathbb{D}^{k+1}.
\end{equation}
Let $\Lambda$ denote
the class of strictly increasing, continuous mappings of $\left[0, \hat T\right]$
onto itself. Then the function space $\mathbb{D}^{k+1}$
is equipped with the Skorohod topology with the metric
\begin{equation}
 \label{eq:skrd}
 d(x,y) =\inf_{\lambda\in\Lambda}\left\{ \left\Vert \lambda-I\right\Vert \vee\left\Vert x\circ\lambda-y\right\Vert \right\} ,\forall x,y\in\mathbb{D}^{k+1}.
\end{equation}

Since $d(x,y) \leq \left\Vert x-y\right\Vert$ the convergence in Skorohod topology does not imply the convergence in uniform topology. However, if the limit is in $\mathbb{C}^{k+1}$ then they are equivalent.

\begin{proposition}
 \label{P:sc}
({\cite[page 124]{Billingsley2009}})
Elements $x_n$ of $\mathbb{D}^{k+1}$ converge to a limit $x$ in the Skorohod topology if and only if there exist functions $\lambda_n$ in $\Lambda$ such that $\lim_n x_n(\lambda_nt)=x(t)$ uniformly in $t$ and $\lim_n \lambda_n t = t$ uniformly in $t$. Moreover, if $x$ is in $\mathbb{C}^{k+1}$, then Skorohod convergence implies uniform convergence.\end{proposition}

\subsubsection{Convergence in distribution} \label{sec:conv}
The notion of the {\it convergence in distribution} (see \cite{Billingsley2009}) or {\it weak convergence} (see \cite{Kushner2001}) may be defined in various ways in different literatures.
Since it plays an important role in Theorem~\ref{t:main} and throughout the paper, we give a short remark for its clarification. The random elements
of our interests $V^{h}$ and $V$ of Theorem~\ref{t:main} are the maps from 
a probability space $(\Omega, \mathcal F, \mathbb P)$ to
a Skorohod metric space $(\mathbb D^{k+1}, d)$ equipped with Skorohod metric 
$d$  defined in \eqref{eq:skrd}, and often write it as
$$V^{h}, V: (\Omega, \mathcal F, \mathbb P) \mapsto (\mathbb D^{k+1}, d)$$
or $V^{h}, V: \Omega \mapsto \mathbb D^{k+1}$  in short if the context has no ambiguity on $\mathcal F, \mathbb P$ and $d$.
The {\it distribution} of the random element $V$ refers to
$\mathbb P V^{-1}$, which is indeed the probability measure 
on the Borel $\sigma$-algebra $\mathcal B(\mathbb D^{k+1})$, i.e.
$$\mathbb P V^{-1} (A) = \mathbb P\{\omega: V(\omega) \in A\}, \ 
\forall A \in \mathcal B(\mathbb D^{k+1}).$$
$\mathbb P V^{-1}$ is sometimes called as the law of $V$, 
or the probability measure induced by $V$, 
or the push forward measure. 
Similarly, $\mathbb P(V^{h})^{-1}$ is the distribution of $V^{h}$.

We say, as $h\to 0$,  
$V^{h}$ converges to $V$ in distribution with respect to 
Skorohod metric $d$, if the measure $\mathbb P (V^{h})^{-1}$ 
weakly converges to $\mathbb PV^{-1}$, denoted by 
$\mathbb P (V^{h})^{-1} \Rightarrow \mathbb PV^{-1}$. 
It sometimes called $V^{h}$ converges to $V$ weakly
with respect to Skorohod metric.
By Portmanteau Theorem, 
one can equivalently define the convergence
in distribution in any of five different ways provided by
Page 26 of \cite{Billingsley2009}. For instance, a common definition adopted in many references is that, $V^{h}$ is said to be convergent to $V$ in distribution with respect to 
Skorohod metric $d$, if $\mathbb E[f(V^{h})] \to \mathbb E[f(V)]$ for all
$f\in C_{b}(\mathbb D^{k+1})$, the space of all
bounded continuous (w.r.t. metric $d$) functions on $\mathbb D^{k+1}$.

One may already note that the convergence in distribution relies on
the topology of $\mathbb D^{k+1}$. Indeed, $C_{b}(\mathbb D^{k+1})$
under uniform topology induced by $\|\cdot \|$ of \eqref{eq:ufm}
 is a bigger space than $C_{b}(\mathbb D^{k+1})$ induced by 
 Skorohod metric $d(\cdot, \cdot)$ of
\eqref{eq:skrd}. This immediately yields by the definition that the 
convergence in distribution with respect to the
uniform topology implies convergence in distribution with respect to
Skorohod topology. In the rest of the paper, unless it is specified, 
the space $\mathbb D^{k+1}$ is
equipped with Skorohod metric $d$ by default, and convergence in 
distribution means by default the convergence in distribution with respect to Skorhod
topology. 

\subsubsection{Discussion on Assumption~\ref{a:nondegenerate}}

Recall the hitting time operator $\pi$ of \eqref{eq:pi01}.
This notion enables us to treat the default time $\tau_{i}$ as a function on a random process, i.e. $\tau_{i} = \pi(V_{i}, L_{i})$. However, one can not
assume $\pi(\cdot, L_{i})$ is continuous in $\mathbb D$ in general
from the following example.
\begin{example}
 \label{e:02}
 $\pi(x, 0)$ is not upper semicontinuous at $x\in \mathbb C$
 given by $$x(t) = |t - 1/ 2 |,$$
 since
 $\lim_{n}\pi(x_{n}, 0) = \hat T > 1/2 = \pi(x, 0)$ where $x_{n} = x+ 1/n$. See Figure~\ref{Fig:tang_prob} for illustration.
 \begin{figure}[hbtp]
\centering
\includegraphics[scale=0.6]{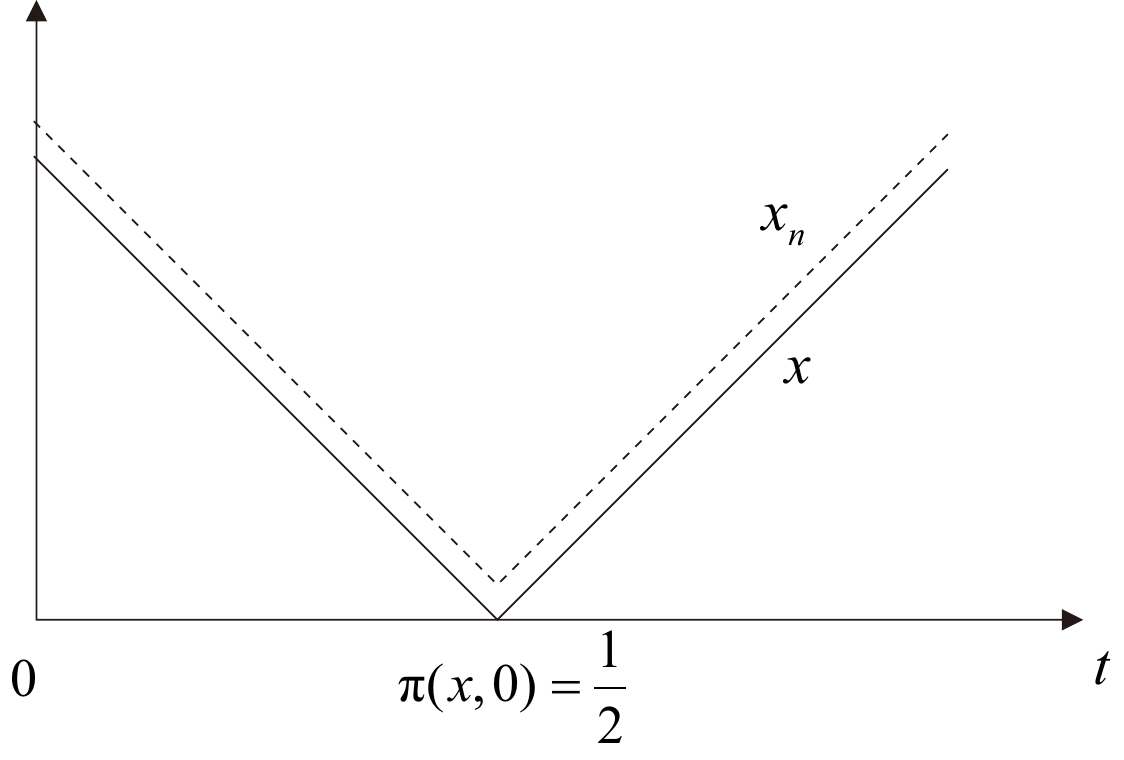}
\caption{Illustration of Example~\ref{e:02}}\label{Fig:tang_prob}
\end{figure}

One can also adapt the above idea to illustrate the potential issue
arising from the 
numerical computation of the credit risk model. Let's assume
that the firm value $V_{1}$ follows the deterministic curve $x+1$, 
i.e. $V_{1}(t) = x(t) +1 = 1+ |t - 1/2|$, and a tradable derivative
price is given by $\hat c = \mathbb E[\tau_{1}]$, where $\tau_{1} = \inf\{t:
V_{1}(t) \le 1\}\wedge \hat T$. One can easily see that an approximation $\hat c^{h}$ 
by usual Euler scheme does not guarantee the convergence $\hat c^{h} \to 
c^{h}$. Obviously, it is not a realistic example due to its zero volatility. Then,
can we avoid mispricing from the computation by assuming non-zero volatility of the firm value?

\end{example}

The nondegenerate condition Assumption~\ref{a:nondegenerate} is very
 important throughout the paper. It not only implies any two of firms
 $V_{i}$ and $V_{j}$ are not perfectly correlated, but also makes the
 barriers regular to the underlying diffusion. As an immediate consequence,
 we have $x\mapsto \pi(x, L_{i})$ continuous $\mathbb PV_{i}^{-1}$-almost surely for all $i$'s.

Another use of Assumption~\ref{a:nondegenerate} is on the total number
of defaults $\alpha$ in Example~\ref{e:01}. One  can  rewrite this $\alpha$ in terms of
$\pi$ by 
\begin{equation}
 \label{eq:alphah}
 \alpha(t) = \sum_{j=1}^{k} I(\pi(V_{i}, L_{i}) \le t) := \hat \alpha (V)
\end{equation}
for some mapping $\hat \alpha$ defined on $\mathbb D^{k+1}$.
Example~\ref{e:02} implies that the functional $\hat \alpha$ above 
is not continuous with respect to Skorohod topology, and hence
it violates the sufficient condition for the weak convergence given
in Condition C5.1 of \cite{Kurtz1991}. However,
Example~\ref{e:03} below can verify the almost sure continuity under 
Assumption~\ref{a:nondegenerate},
which is enough for our purpose.

\begin{example}\label{e:03}
The number of defaults $\alpha$ defined
in Example~\ref{e:01} can be rewritten by \eqref{eq:alphah}.
Example~\ref{e:02} implies that $\hat \alpha$ is not continuous in
general. However, under Assumption~\ref{a:nondegenerate}, $\hat\alpha$
is continuous under Skorohod topology almost surely with respect to
$\mathbb P V^{-1}$. Indeed, this follows from the following two facts:
\begin{enumerate}
 \item By Blumenthal 0-1 law (see \cite{Durrett2010}),
 $\pi(\cdot, L_{i})$ is continuous
 with respect to Skorohod metric almost surely in $\mathbb PV_{i}^{-1}$
 for all $i= 0, \ldots, k$, i.e.
 $$\mathbb P(V_{i} \in \{x\in \mathbb D: \pi(\cdot, L_{i}) \hbox{ is continuous at } x\}) = 1.$$
 \item
 Moreover,
 $I(\pi(\cdot, L_{j})\le t)$ is also continuous
 with respect to Skorohod metric almost surely in $\mathbb PV_{j}^{-1}$,
 the probability induced by $V_{j}$.
\end{enumerate}
\end{example}

\subsubsection{Discussion on Assumption~\ref{a:cont_phi}}
\begin{figure}[hbtp]
\centering
\includegraphics[scale=0.5]{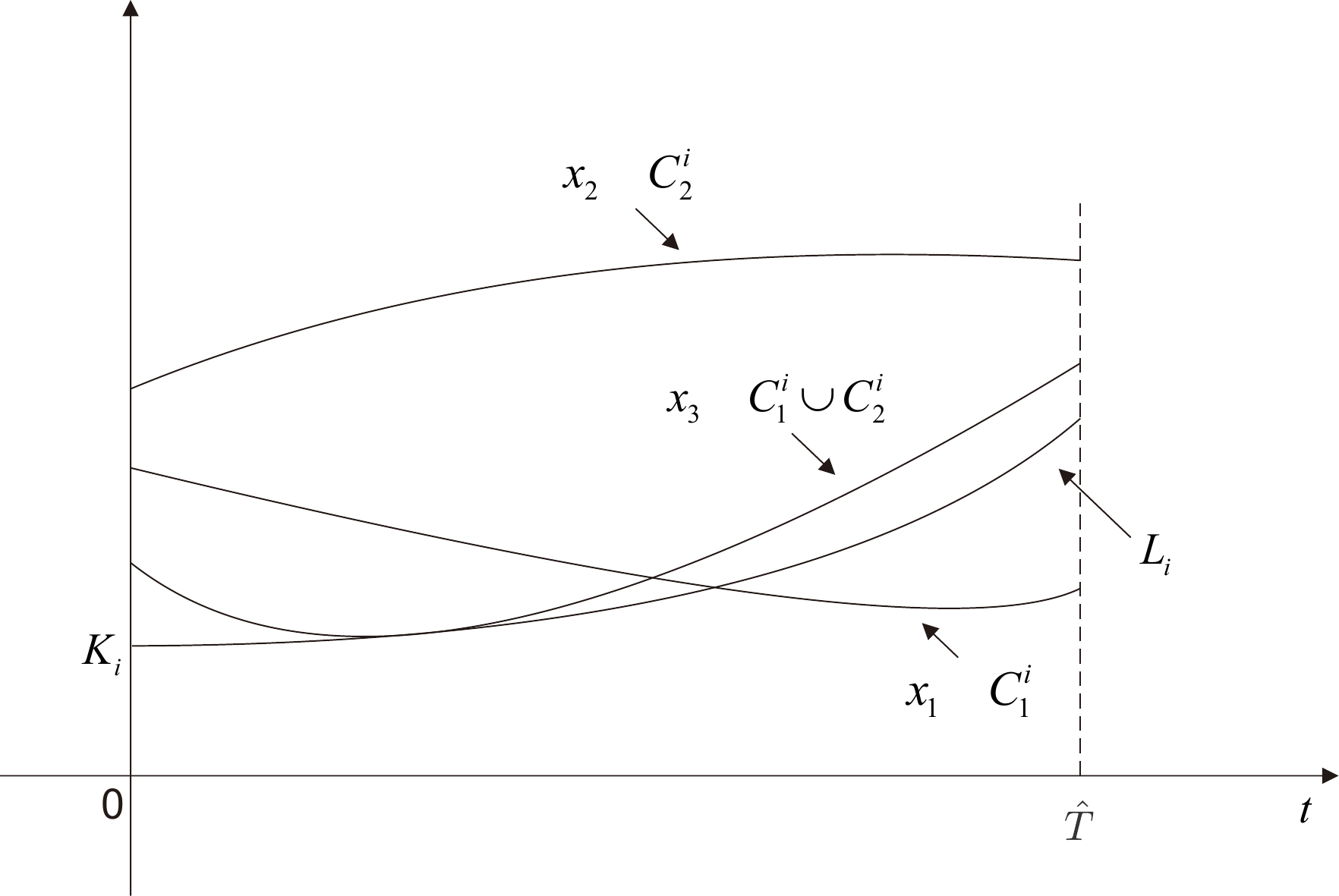}
\caption{Illustration of disjoint subsets:
$C_{1}^{i}$ and $C_{2}^{i}$. $x_{1}$ and $x_{2}$ belong to $C_{1}^{i}$
and $C_{2}^{i}$, respectively. $x_{3}$ doesn't belong to $C_{1}^{i}$
or $C_{2}^{i}$.}\label{Fig:C1&C2}
\end{figure}

Condition C5.1 in \cite{Kurtz1991} requires the mapping $x\mapsto \sigma(x,\cdot)$ is continuous under Skorohod topology in
$\D^{k+1}$, while our volatility violates this condition as of Example~\ref{e:03}.
In contrast, Assumption~\ref{a:cont_phi} may be regarded as a
continuity requirement of coefficient functions on a smaller space
$\{x\in \C^{k+1}: x_i\in C_1^i \cup C_2^i,i=0,1,\dots,k\},$
Figure~\ref{Fig:C1&C2} provides some examples to illustrate the concept of $C_{1}^{i}$
and $C_{2}^{i}$. The paths in the union set of $C_1^i$ and $C_2^i$ are regular with respect to the boundary $L_i$ in the sense that once the path touches the boundary $L_i$ the path pushes through the boundary.
Continued from Example~\ref{e:03}, due to the fact $\mathbb P(V_{i}\in C_{1}^{i} \cup C_{2}^{i}) = 1$ under Assumption~\ref{a:nondegenerate},
$\alpha$ also satisfies Assumption~\ref{a:cont_phi}. We refer Section 2.2
of \cite{Song2012} for more detailed descriptions.

\subsubsection{Discussion on Assumption~\ref{a:zero_corr} or \ref{a:ps_const}}
Assumption \ref{a:zero_corr} or \ref{a:ps_const}, together with Assumption~\ref{a:nondegenerate} is to ensure the indicator function
$\mathbf{1}\left(\tau_{0}>\tau_{\left(i\right)}\wedge T\right)$ is
continuous $\mathbb{P}V^{-1}$-almost surely. Suppose the CDS writer is default free before maturity
$T$, then the indicator function $\mathbf{1}\left(\tau_{0}>\tau_{\left(i\right)}\wedge T\right)$
is a constant one a.s. $\mathbb{P}V^{-1}$. For such a case, Assumption \ref{a:zero_corr} or \ref{a:ps_const} is not required in Theorem~\ref{t:main}. From the financial perspective, Assumption~\ref{a:zero_corr} implies that CDS writer and the reference names are independent of each other from exogenous factors ($W_0$ and $W_i$ are independent). This is realistic in practice since an important  criteria adopted by practitioners for choosing an appropriate CDS writer is that the CDS writer has little correlation with the reference names.  On the other hand,
if the CDS writer has non-zero correlation with reference names (but not
perfectly correlated due to Assumption~\ref{a:nondegenerate}, then
one can still have almost sure continuity of $\mathbf{1}\left(\tau_{0}>\tau_{\left(i\right)}\wedge T\right)$ by requiring
the piecewise constant form of $\sigma$, which is indeed the case in reality.
Note that the volatility is calibrated in practice form time to time, not continuously.

\subsection{Other Operators Related to the CDS Pricing Formula}
For the later use, we also introduce other related notions here.
For any $n\in\mathbb{N}$, define $\mathcal{S}^{n}:\mathbb{R}^{n}\times\left\{ 1,2,\dots,n\right\} \mapsto\mathbb{R}$
as
\[
\mathcal{S}^{n}\left(x,j\right)
:=\text{the $j$th smallest value among }\left\{ x_{i}\right\} _{i=1}^{n}.
\]
Finally, define $F_{i}:\mathbb{D}^{k+1}\mapsto\mathbb{R}$, $i=1,2$, as
\begin{align*}
F_{1}\left(x\right) & :=e^{-r\mathcal{S}^{k}\left(\left\{ \pi\left(x_{n},L_{n}\right)\right\} _{n=1}^{k},i\right)}\left(1-\delta_{i}\right)\mathbf{1}\left(\mathcal{S}^{k}\left(\left\{ \pi\left(x_{n},L_{n}\right)\right\} _{n=1}^{k},i\right)\leq T\right)
\\
 & \quad\mathbf{1}\left(\pi\left(x_{0},L_{0}\right)>\mathcal{S}^{k}\left(\left\{ \pi\left(x_{n},L_{n}\right)\right\} _{n=1}^{k},i\right)\wedge T\right),
\end{align*}
and
\begin{align*}
F_{2}\left(x\right) & :=\sum_{j=1}^{m}e^{-rt_{j}}\Delta t_{j}\mathbf{1}\left(\mathcal{S}^{k}\left(\left\{ \pi\left(x_{n},L_{n}\right)\right\} _{n=1}^{k},i\right)>t_{j}\right)\mathbf{1}\left(\pi\left(x_{0},L_{0}\right)>t_{j}\right).
\end{align*}

With the help of mappings $\pi$, $\mathcal S^k$ and $F_i$, $i=1,2$, the default time $\tau_i$ of  company $i$, the $i$th default time $\tau_{(i)}$ of the reference portfolio, and the swap rates 
$\hat{c}_i$
and 
$\hat{c}_i^h$
can be expressed respectively as follows:
\begin{align*}
\tau_i &= \pi(V_i,L_i),\quad i=0,1,\dots,k,\\
\tau_{(i)} &=S^k(\{\pi(V_j,L_j)\}_{j=1}^k,i),\quad i=1,2,\dots,k,\\
\hat{c}_i
&=\frac{\mathbb{E}[F_1(V)]}{\mathbb{E}[F_2(V)]}\quad \text{and}\quad 
\hat{c}_i^h
=\frac{\mathbb{E}[F_1(V^h)]}{\mathbb{E}[F_2(V^h)]}.
\end{align*}

\begin{figure}[hbtp]

\centering
\includegraphics[scale=0.5]{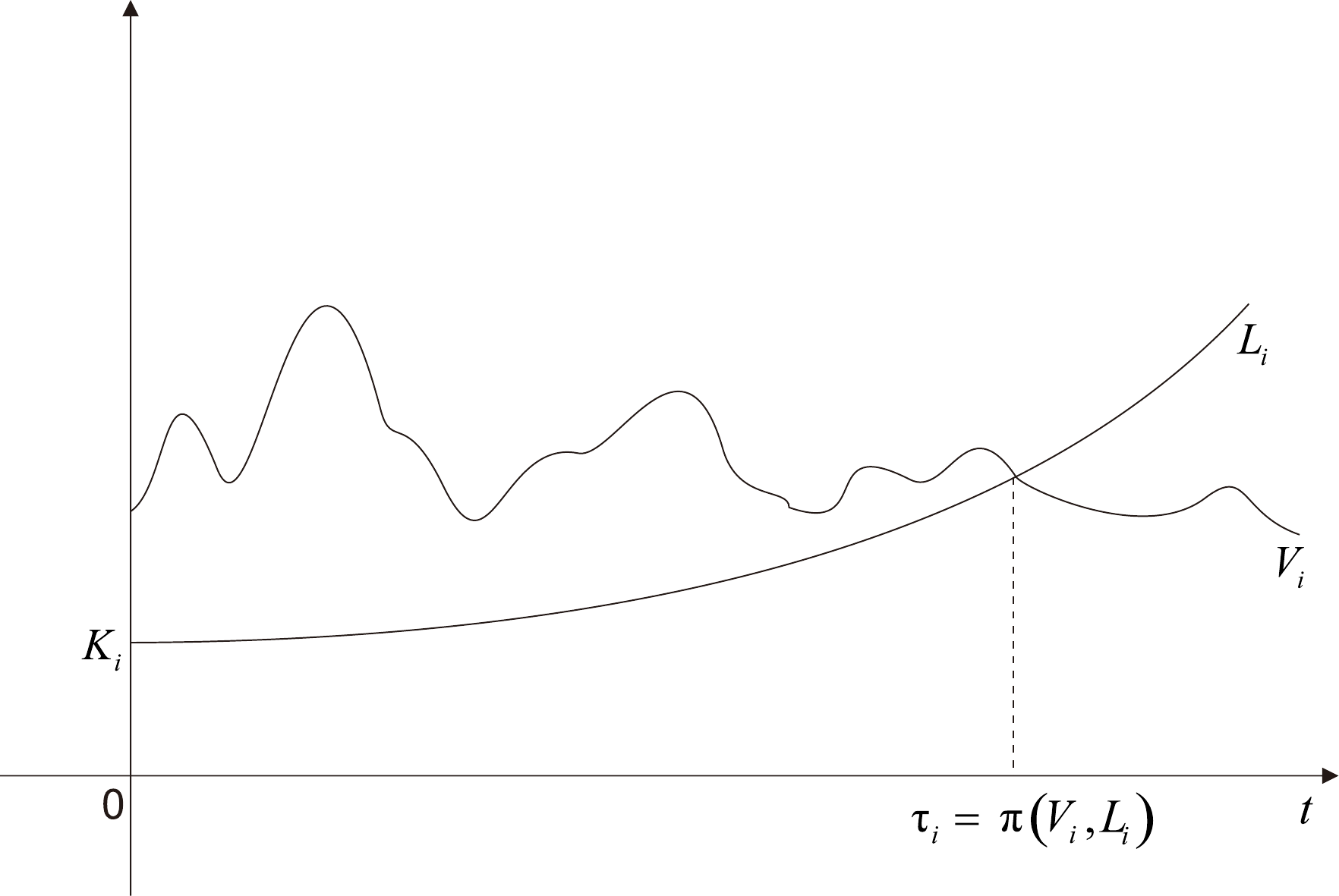}
\caption{Illustration of the default time for  company $i$, $\tau_{i}=\pi\left(V_{i},L_{i}\right)$.}\label{Fig:tau_i}
\end{figure}

The default time for  company $i$ is illustrated in Figure~\ref{Fig:tau_i}.

\section{Proof of Theorem~\ref{t:main}}
In this section we prove Theorem \ref{t:main} through a number of lemmas.

\subsection{Preliminary Estimates}

We discuss some properties of $V^{h}$ here, which plays a
crucial role in the subsequent parts. Let $z^h(t):=[\frac{t}{h}]h$. The Euler scheme can be rewritten in terms of interpolated process in the
following way:
$$V^{h}(t) = V^{h}_{[\frac{t}{h}]} = V_{0} + \int_{0}^{z^{h}(t)}
diag(V^{h}(s))\mu^{h}(s) ds + \int_{0}^{z^{h}(t)} diag(V^{h}(s))\sigma^{h}(s) dW(s).$$ For convenience, we also denote
$\Delta V_{n}^{h}:=V_{n+1}^{h}-V_{n}^{h}$ and
$\Delta M_{n}^{h}:=\Delta V_{n}^{h}-E\left[\Delta V_{n}^{h}\vert\mathcal{F}_{n}^{h}\right].$

\begin{lemma}
 \label{l:mod1}
 If Assumption~\ref{a:bs} holds, then $V^{h}$ satisfies, for $p\geq 2$ and  $h>0$,
\begin{equation}\label{eq:mod12}
 \mathbb E\left[\sup_{0\leq t\leq \hat T}\left|V^{h}(t)\right|^{p}\right] \le K,
\end{equation}
\begin{equation} \label{eq:mod11}
 \mathbb E\left[ \sup_{[\frac{t_1}{h}]\leq m\leq [\frac{t_2}{h}]-1}\left\vert \sum_{n=[\frac{t_1}{h}]}^m \Delta V_n^h\right\vert^p \right]\leq K\left[z^h(t_2)-z^h(t_1)\right]<K(t_2-t_1+h),\quad \forall t_1<t_2,
\end{equation}
\begin{equation} \label{eq:mod13}
 \mathbb E\left[ \sup_{[\frac{t_1}{h}]\leq m\leq [\frac{t_2}{h}]-1}\left\vert \sum_{n=[\frac{t_1}{h}]}^m \Delta V_n^h\right\vert^p  \bigg| \mathcal{F}_{t_1}  \right]\leq K\E\left[ \int_{z^h(t_1)}^{z^h(t_2)}|V^h(s)|^p ds \bigg| \mathcal{F}_{t_1} \right],\quad \forall t_1<t_2,
\end{equation}
where $[x]$ is the largest integer not greater than $x$.
\end{lemma}
\textit{Proof:} For any $t\in[h,\hat T]$, $h>0$ and $p\geq 2$, using Burkholder inequality and Assumption~\ref{a:bs} we estimate $p$th moment of $V^h(t)$ as follows:
\begin{align*}
\E|V^h(t)|^p \leq & \E (|V_0|+|\int_0^{z^h(t)}diag(V^h(s))\mu^h(s)ds|+|\int_0^{z^h(t)}diag(V^h(s))\sigma^h(s)dW(s)|)^p\\
\leq & K\E (|V_0|^p+|\int_0^{z^h(t)}diag(V^h(s))\mu^h(s)ds|^p+|\int_0^{z^h(t)}diag(V^h(s))\sigma^h(s)dW(s)|^p)\\
\leq & K+\E \int_0^{z^h(t)}|diag(V^h(s))\mu^h(s)|^p ds +K\E[\int_0^{z^h(t)}|diag(V^h(s))\sigma^h(s)\sigma^h(s)^T diag(V^h(s))|ds]^{p/2}\\
\leq & K+K\int_0^t \E|V^h(s)|^p ds.
\end{align*}
From above inequality and Gronwall's inequality, we have
\begin{equation}
\label{eq:E_V^h^4}
\E|V^h(t)|^p \leq Ke^K.
\end{equation}
For the case of $t<h$, \eqref{eq:E_V^h^4} still holds by noting $V^h(t)=V^h(0)$ when $0\leq t<h$.
For any $0\leq t_1<t_2\leq \hat T$, we have
\begin{align*}
&\sup_{[\frac{t_1}{h}]\leq m\leq [\frac{t_2}{h}]-1}\left|\sum_{n=[\frac{t_1}{h}]}^{m}\Delta V_{n}^{h}\right|^{p}\\
& = \sup_{[\frac{t_1}{h}]\leq m\leq [\frac{t_2}{h}]-1}\left| \int_{z^h(t_1)}^{(m+1)h}diag(V^h(s))\mu^h(s)ds + \int_{z^h(t_1)}^{(m+1)h}diag(V^h(s))\sigma^h(s)dW(s)\right|^{p} \\
& \leq  K\sup_{[\frac{t_1}{h}]\leq m\leq [\frac{t_2}{h}]-1}\left| \int_{z^h(t_1)}^{(m+1)h}diag(V^h(s))\mu^h(s)ds \right|^p + K\sup_{[\frac{t_1}{h}]\leq m\leq [\frac{t_2}{h}]-1} \left| \int_{z^h(t_1)}^{(m+1)h}diag(V^h(s))\sigma^h(s)dW(s)\right|^{p}\\
& \leq  K \int_{z^h(t_1)}^{z^h(t_2)}\left|diag(V^h(s))\mu^h(s)\right|^p ds  + K\sup_{[\frac{t_1}{h}]\leq m\leq [\frac{t_2}{h}]-1} \left| \int_{z^h(t_1)}^{(m+1)h}diag(V^h(s))\sigma^h(s)dW(s)\right|^{p}.
\end{align*}
Since the integral of the second term is a martingale, applying \eqref{eq:E_V^h^4}, Assumption~\ref{a:bs}, Burkholder's inequality and Holder inequality gives us
\begin{align*}
\E \sup_{[\frac{t_1}{h}]\leq m\leq [\frac{t_2}{h}]-1}\left|\sum_{n=[\frac{t_1}{h}]}^{m}\Delta V_{n}^{h}\right|^{p}
 \leq &  K\E \int_{z^h(t_1)}^{z^h(t_2)}\left|diag(V^h(s))\right|^p ds  + K\E \bigg[ \int_{z^h(t_1)}^{z^h(t_2)}\left|diag(V^h(s))\sigma^h(s)\right|^2 ds\bigg]^{p/2} \\
 \leq &  K\E \int_{z^h(t_1)}^{z^h(t_2)}\left|diag(V^h(s))\right|^p ds\label{eq:Exp_sup_dV}\numberthis\\
\leq & K(t_2-t_1+h).
\end{align*}
Setting $t_1 =0$ and $t_2 = \hat T$, we obtain
\begin{align*}
\E \bigg[\sup_{0\leq t\leq \hat T}\big|V^h(t)\big|^p \bigg ] \leq & K|V_0|^p + K\E \bigg[\sup_{0\leq m\leq [\frac{\hat T}{h}]-1}\bigg|\sum_{n=0}^m \Delta V^h_n\bigg|^p\bigg]\\
\leq & K.
\end{align*}
A review of the proof shows that the inequality \eqref{eq:Exp_sup_dV} still holds when the expectation is replaced by the conditional expectation. Hence we have \eqref{eq:mod13}.
\hfill$\square$


\begin{lemma}
 \label{le:jump}
 If Assumption~\ref{a:bs} holds, then $j(V^{h}) \Rightarrow 0$ as $h\to 0$, where $j(x) := \sup_{0< t \le T} |x(t) - x(t^-)|$ for $x\in \mathbb D^{k+1}$. 
\end{lemma}
\textit{Proof:} Since boundedness of $\mu^h$ and $\sigma^h$ implies that
 $$|\Delta V_{n}^{h}| \le K |V_{n}^{h}| (h + \sqrt h |Z_{n+1}|),$$
we have
$$
j(V^h)=\sup_n\vert\Delta V_n^h\vert\leq \underbrace{Kh\sup_n \vert V_n^h\vert}_{\text{I$^h$}}+\underbrace{K\sqrt{h}\sup_n \vert V_n^h\vert\sup_n \vert Z_{n+1}\vert}_{\text{II$^h$}},
$$
where in above inequality the first and second term are denoted by I$^h$ and II$^h$, respectively.

\noindent The convergence of $E[\text{I$^h$}]$ to zero as $h\rightarrow 0$ can be obtained from 
$$\mathbb E[\text{I$^h$}] \le K h \mathbb E [ \sup_{n} |V_{n}^{h}|] \le K h (
\mathbb E [ \sup_{n} |V_{n}^{h}|^{2}])^{1/2} \le Kh.$$
The last inequality in the above is due to \eqref{eq:mod12}. On the other hand,
\begin{eqnarray*}
\mathbb E[\text{II}^h]&\leq&K\sqrt{h}\left(
\mathbb E\sup_n\vert V^h_n\vert^{4\over 3}\right)^{3\over 4} \left(
\mathbb E\sup_n\vert Z_{n+1}\vert^{4}\right)^{1\over 4}\\
&=&Kh^{1\over 4}\left(
\mathbb E\sup_n\vert V^h_n\vert^{4\over 3}\right)^{3\over 4} \left(h 
\mathbb E\sup_n\vert Z_{n+1}\vert^{4}\right)^{1\over 4}\\
&\leq&Kh^{1\over 4}\left(
\mathbb E\sup_n\vert V^h_n\vert^{4\over 3}\right)^{3\over 4} \left({ 
\mathbb E\sum_{n=1}^N\vert Z_{n}\vert^{4}\over N}\right)^{1\over 4}\\
&=&Kh^{1\over 4}\left(
\mathbb E\sup_n\vert V^h_n\vert^{4\over 3}\right)^{3\over 4} \left(
\mathbb E\vert Z_{1}\vert^{4}\right)^{1\over 4}\\
&\rightarrow&0\text{\quad as }h\rightarrow 0
\end{eqnarray*}
The conclusion follows from that $j(V^h)$ converges to 0 in $L^1$ as $h\rightarrow 0$.\hfill$\square$

\begin{remark}
\label{Re:Skorohod_top_C}
Lemma \ref{le:jump} is the key result that enables us to show that the weak limit
process $\bar V$ of $\{V^{h}\}$ is continuous.
Skorohod representation theorem allows us to treat the limit $V^{h} \Rightarrow \bar V$ in almost
sure sense, that is
\begin{equation} \label{eq:lim1}
\lim_{h} d(V^{h}, \bar V) = 0, \ a.s.
\end{equation}
However, it does not imply almost sure limit with uniform topology, i.e.
\begin{equation}
 \label{eq:lim2}
 \lim_{h} \|V^{h} - \bar V\| = 0, \  a.s.
\end{equation}
 may not be true. In other words, \eqref{eq:lim1} does not imply
$$\lim_{h} f(V^{h}(t)) = f(\bar V(t)), \ \forall t$$ even for a bounded continuous function $f$, which is useful in characterizing
properties of $\bar V$.
Indeed, one shall prove continuity of $\bar V$ (i.e. $\mathbb P\{\bar V\in
\mathbb C^{k+1}\} = 1$) in advance to make use of \eqref{eq:lim2} from Proposition~\ref{P:sc}. We note that the proof on the continuity
of $\bar V$ is missing in \cite{Song2013} and the related references therein.
\end{remark}

\subsection{Weak Convergence of Approximating Solutions}
This part shows that the approximating processes $V^{h}$ converge to
$V$ in distribution. A general approach to this goal is first to prove tightness of $\{\mathbb P (V^{h})^{-1}\}$ for extracting a weak limit from any subsequence, then to
apply the Skorohod representation theorem for passing the limit
almost surely, and finally to characterize the limiting process as the solution of the underlying SDE, provided there exists a unique weak solution.  The main result of this subsection is Theorem~\ref{thm:V_wk_conv}.

\begin{proposition}
\label{prop:UE_weak} If Assumption~\ref{a:bs} holds, then
there exists a unique weak solution to SDE \eqref{eq:dVt}.
\end{proposition}
\textit{Proof:} For the sake of simple presentation, we assume no jump for $\mu$. Taking logarithm, it is sufficient to consider the following SDE:
\begin{equation}
 \label{eq:log_dVt}
dV\left(t\right)=\mu\left(V,t\right)\thinspace dt+\sigma\left(V,t\right)\thinspace dW\left(t\right),\quad t\geq 0.
\end{equation}
By definition, $V$ can be constructed uniquely by the following steps:
\begin{enumerate}
\item Let $V_1(t) = V_0 +\int_0^t\mu(s) ds + \sigma_c(s) dW(s)$ for $t>0$.
\item Let $\tau_1=\inf\{t>0:\mathcal{N}^\sigma(V_1,t)=1\}\wedge \hat T$.
\item Let $V_2(t) = V_1(t)$ for $t\leq \tau_1$, otherwise,
\[
V_2(t)=V_1(\tau_1)+\int_{\tau_1}^t \mu(s) ds +(\sigma_c(s)+J_1^{\sigma}(V_1))dW(s)
\]
for $t>\tau_1$.
\item Let $\tau_2=\inf\{t>\tau_1:\mathcal{N}^{\sigma}(V_2,t)=2\}\wedge \hat T$.
\item Repeat above steps to construct $V_i$, $\tau_i$, $i=1,2,\dots$ until $\tau_i= \hat T$.
\end{enumerate}
According to \cite[Theorem 1.6.3]{Yong1999}, $V_1$ has a unique strong solution from 0 to $\hat T$, hence $\tau_1$ is well defined. Since $\tau_1$ is no greater than $\hat T$, $V_1$ has a unique solution from 0 to $\tau_1$. \cite[Theorem 1.6.3]{Yong1999} also implies that $\E |V_1(\tau_1)|^2\leq K$, so $V_2$ has a unique strong solution from $\tau_1$ to $T$ with initial value $V_1(\tau_1)$ at time $\tau_1$. Therefore, $V_2$ has a unique strong solution from 0 to $\hat T$. Since the number of jumps of $\sigma$ is finite due to Assumption~\ref{a:bs}, we can proceed the above procedure inductively until $\tau_n=\hat T$, where $n$ is some finite number. Then $V_n$ is the unique strong solution of \eqref{eq:log_dVt} from 0 to $\hat T$. Hence, the uniqueness of weak solution is ensured according to \cite[Theorem 9.1.7]{Revuz1999}.
\hfill $\square$
\begin{lemma}
\label{le:tight_V^h} If Assumption~\ref{a:bs} holds, then $\{\P(V^h)^{-1}:h>0\}$ is tight.
\end{lemma}
\textit{Proof:} According to \cite[Theorem 3.8.6]{Ethier2005} it is sufficient to verify condition a) in \cite[Theorem 3.7.2]{Ethier2005} and condition b) in \cite[Theorem 3.8.6]{Ethier2005} in order to show $\{\P(V^h)^{-1}:h>0\}$ is tight. To verify condition a) in \cite[Theorem 3.7.2]{Ethier2005} we need to verify that for every $\epsilon$ and rational $t\geq 0$, there exists a compact set $\Gamma_{\epsilon,t}\subset\R^{k+1}$ such that
\[
\inf_{h}\P\{V^h(t)\in \Gamma_{\epsilon,t}^\epsilon\}\geq 1-\epsilon,
\]
where $\Gamma_{\epsilon,t}^\epsilon:=\{x\in\R^{k+1}:\inf_{y\in \Gamma_{\epsilon,t}}|x-y|<\epsilon\}$.
It is worth noting that $\Gamma_{\epsilon,t}^\epsilon\supset \Gamma_{\epsilon,t}$. Denote by $K_0$  the bound of $\E|V^h(t)|^4$ from \eqref{eq:mod12}. For every $\epsilon$ and $t\geq 0$, by setting $\delta = (\frac{K_0}{\epsilon})^{\frac{1}{4}}$ and $\Gamma_{\epsilon,t}=\{x\in \R^{k+1}:\vert x\vert\leq \delta \}$, we have
\begin{align*}
\inf_{h>0}\mathbb{P}(V^h(t)\in \Gamma_{\epsilon,t}^\epsilon)= & 1-\sup_{h>0}\mathbb{P}(V^h(t)\notin \Gamma_{\epsilon,t}^\epsilon)\\
\geq & 1-\sup_{h>0}\mathbb{P}(V^h(t)\notin \Gamma_{\epsilon,t})\\
= & 1-\sup_{h>0}\mathbb{P}(|V^h(t)|>\delta)\\
\geq & 1- \sup_{h>0}\frac{\E\vert V^h(t)\vert^4}{\delta^4} = 1-\epsilon.
\end{align*}
To verify condition b) in \cite[Theorem 3.8.6]{Ethier2005}, we need to find some positive $\beta$ and a family $\{\gamma^h(\delta):0<\delta<1,\mbox{ all $h$}\}$ of  nonnegative random variables  satisfying
\begin{equation}
\label{eq:tight_cond_1}
\E [|V^h(t+u)-V^h(t)|^\beta \wedge 1|\mathcal{F}_t](|V^h(t)-V^h(t-\nu)|^\beta\wedge 1)\leq \E[\gamma^h(\delta)|\mathcal{F}_t]
\end{equation}
for $0\leq t\leq T$, $0\leq u\leq \delta$, and $0\leq v\leq \delta\wedge t$; in addition,
\begin{equation}
\label{eq:tight_cond_2}
\lim_{\delta\to 0}\sup_h \E[\gamma^h(s)]=0
\end{equation}
and
\begin{equation}
\label{eq:tight_cond_3}
\lim_{\delta\to 0}\sup_h \E[|V^h(\delta)-V^h(0)|^\beta \wedge 1]=0.
\end{equation}
We claim that $\beta =4$ and a family of nonnegative random variables $\{\sup_{0\leq s\leq T}|V^h(s)|^4[\delta+2(h\wedge2\delta)]:\delta>0,h>0\}$ satisfy \eqref{eq:tight_cond_1}, \eqref{eq:tight_cond_2} and \eqref{eq:tight_cond_3}. For \eqref{eq:tight_cond_1}, since either $|V^h(t+u)-V^h(t)|^4$ or $|V^h(t)-V^h(t-\nu)|^4$ is zero when $h>2\delta$ due to piecewise constant form of $V^h$, we then have
\begin{align*}
& \E [|V^h(t+u)-V^h(t)|^4 \wedge 1|\mathcal{F}_t](|V^h(t)-V^h(t-\nu)|^4\wedge 1)\\
= & \E [|V^h(t+u)-V^h(t)|^4 \wedge 1|\mathcal{F}_t](|V^h(t)-V^h(t-\nu)|^4\wedge 1)\mathbf{1}(h\leq 2\delta)\\
\leq & \E [\sup_{[\frac{t}{h}]\leq m\leq [\frac{t+\delta+h}{h}]-1}\bigg|\sum_{n=[\frac{t}{h}]}^m \Delta V_n^h\bigg|^4\big|\mathcal{F}_t]\mathbf{1}(h\leq 2\delta)\\
\leq & \E [\int_{z^h(t)}^{z^h(t+\delta+h)}|V^h(s)|^4ds\big|\mathcal{F}_t]\mathbf{1}(h\leq 2\delta)\\
\leq & \E [\sup_{0\leq s\leq T}|V^h(s)|^4[\delta+2(h\wedge2\delta)]\big|\mathcal{F}_t].
\end{align*}
\eqref{eq:tight_cond_2} follows from \eqref{eq:mod12} and Cauchy Schwartz inequality that
\[
\lim_{\delta \to 0}\sup_h \E[\sup_{0\leq s\leq T}|V^h(s)|^4[\delta+2(h\wedge2\delta)]]\leq \lim_{\delta \to 0}K\delta.
\]
\eqref{eq:tight_cond_3} is shown as follows
\begin{align*}
\lim_{\delta \to 0}\sup_{h>0} \E[|V^h(\delta)-V^h(0)|^4]=&\lim_{\delta \to 0}\sup_{0<h\leq \delta}\E[|V^h(\delta)-V^h(0)|^4]\\
\leq & \lim_{\delta \to 0}\sup_{0<h\leq \delta}\E[\sup_{0\leq m\leq [\frac{\delta+h}{h}]-1}|\sum_{n=0}^m \Delta V_n^h|^4] \\
\leq & \lim_{\delta \to 0}\sup_{0<h\leq \delta}K(\delta + h +h) =0. \mbox{\quad (by \eqref{eq:mod11})}
\end{align*}
So, $\{\P(V^h)^{-1}:h>0\}$ is tight.
\hfill $\square$


The next result is needed in the proof of Theorem \ref{thm:V_wk_conv}.
\begin{lemma}
\label{le:Ros_ineq}
(Rosenthal's inequality, \cite{Rosenthal1970})
Let $p\geq 2$ and $(X_i)_{i\in \N}$ be a sequence of independent random variables such that, for any $n\in\N$ and any $i\in\{1,2,\dots,n\}$, $\E(X_i)=0$ and $\E(|X_i|^p)<\infty$. Then we have
\[
\E(|\sum_{i=1}^n X_i|^p)\leq c_p \max(\sum_{i=1}^n\E|X_i|^p,(\sum_{i=1}^n\E (X_i^2))^{p/2}),
\]
where $c_p$ is some constant depending on $p$.
\end{lemma}

We  now state the main theorem of this subsection on weak convergence of the Euler scheme.

\begin{theorem}
\label{thm:V_wk_conv}
Let $V^{h}$ be the Euler approximating
process for $V$. If Assumptions~\ref{a:bs}, \ref{a:nondegenerate} and \ref{a:cont_phi} hold, then $V^{h}$ converges weakly to $V$ as $h\to0$.
\end{theorem}

\noindent
\textit{Proof}: Since $\{\P(V^h)^{-1}:h>0\}$ is tight, hence for an arbitrary infinite sequence, there exists a sub-sequence that has a weak limit. We denote this sub-sequence again by $\left\{ V^h\right\} $, and
its limit by $\overline{V}$. Due to the uniqueness
of weak solution (see Proposition \ref{prop:UE_weak}), it suffices to show that $\overline{V}$ is the
weak solution of \eqref{eq:dVt}. Tightness of $\{\mathbb{P}(V^h)^{-1}:h>0\}$ implies $\overline{V}$ is in $\mathbb{D}^{k+1}$. Moreover, since $j(V^h)\Rightarrow 0$ from Lemma~\ref{le:jump}, $\overline{V}$ is a continuous process due to \cite[Theorem 13.4]{Billingsley2009}. Using Skorohod representation (see \cite[Theorem 9.17]{Kushner2001}) we can find $\widetilde{V}^h$ and $\hat{V}$ in the same probability space $(\widetilde{\Omega},\widetilde{\mathcal{F}},\{\widetilde{\mathcal{F}}_t\},\widetilde{\mathbb{P}} )$ such that their distributions are the same as those of $V^h$ and $\overline{V}$, respectively, and $\widetilde{V}^h$ converges to $\hat{V}$ almost surely. According to Remark~\ref{Re:Skorohod_top_C} and Proposition~\ref{P:sc}, $\widetilde{V}^h$ converges to $\hat{V}$  a.s. under both Skorohod topology and uniform topology. Since $\sigma^h$ is uniformly bounded from below by a positive constant due to Assumption~\ref{a:nondegenerate}, $\hat{V}_i,i=0,1,\dots,k$, is regular with respect to $L_i$ (see \cite[Proposition A.1]{Song2012}). Hence $\phi(\widetilde{V}^h,\cdot) \to \phi(\hat{V},\cdot)$ a.s.  for $\phi=\mu,\sigma$ in Skorohod topology as $h\to 0$ by Assumption~\ref{a:cont_phi}. Denote by $D(\hat{V},\phi)$  the set $\{t\in [0,T]: \phi(\hat{V},\cdot) \mbox{ is discontinuous at } t \}$. Since the number of discontinuities of $\phi(\hat{V},t)$ is bounded almost surely, we can write $[0,T]\setminus D(\hat{V},\phi)$ as the finite union of disjoint  intervals $I_i$, $i=1,\ldots,K$. Since $\phi(\hat{V},\cdot)$ is Holder-1/2 continuous at each $I_i$, $\phi(\hat{V},\cdot)$ is uniformly continuous at each $I_i$. Therefore as $h\to 0$,
\[
\parallel \phi(\widetilde{V}^h,\cdot)-\phi(\hat{V},\cdot)\parallel_{I_i}\to 0, \mbox{$\quad$ a.s., $1\leq i\leq K.$}\label{eq:phi_uniform_conv}\numberthis
\]
For $t\in [0,T]$ denote by
\[
M\left(V, t\right):=V\left(t\right)-V\left(0\right)-\int_{0}^{t}diag\left(V\left(s\right)\right)\mu\left(V,s\right)\thinspace ds.
\]
Noting that $D(\hat{V},\phi)$ is finitely many by Assumption~\ref{a:bs} and $\hat{V}(0)=V^h(0)=V_0$, we have
\begin{align*}
&\lim_{h\to 0}|M(\widetilde V^h, t)-M(\hat V, t)|\\
=&\lim_{h\to 0}| \widetilde{V}^h(t)-\hat{V}(t)|
+\lim_{h\to 0}\bigg|\sum_{i=1}^K\int_{I_i\cap [0,t]}diag(\widetilde{V}^h(s))\mu(\widetilde{V}^h,s)-diag(\hat{V}(s))\mu(\hat{V},s)ds\bigg|\\
\leq &\lim_{h\to 0}\bigg|\int_{[0,t]}diag(\widetilde{V}^h(s)-\hat{V}(s))\mu(\widetilde{V}^h,s)ds\bigg| +\lim_{h\to 0}\bigg|\sum_{i=1}^K\int_{I_i\cap [0,t]}diag(\hat{V}(s))[\mu(\widetilde{V}^h,s)-\mu(\hat{V},s)]ds\bigg|\\
\leq & K\lim_{h\to 0}\parallel \widetilde{V}^h-\hat{V}\parallel_{[0,t]}+\lim_{h\to 0}\sum_{i=1}^K\parallel \mu(\widetilde{V}^h,\cdot)-\mu(\hat{V},\cdot)\parallel_{I_i\cap [0,t]}\int_{I_i\cap [0,t]} diag(\hat{V}(s))ds\\
=&0.
\end{align*}
Moreover, noting that $\widetilde{V}^h$ and $V^h$ have the same law and $\{M(\widetilde{V}^h,t):h>0\}$ is uniformly integrable due to \eqref{eq:mod12}, we have
\[
\E[M(\overline{V}, t)]=\E[M(\hat V, t)]=\lim_{h\to 0}\E[M(\widetilde{V}^h, t)]=\lim_{h\to 0}\E[M({V}^h, t)],
\]
which shows that
\begin{align*}
\E[M(\overline{V}, t)]=&\lim_{h\to0}\mathbb{E}\biggl[\sum_{n=1}^{[\frac{t}{h}]}\biggl(\left(V_{n}^{h}-V_{n-1}^{h}\right)-\int_{\left(n-1\right)h}^{nh}diag\left(V_{n-1}^{h}\right)\mu(V^h,s)\thinspace ds\biggr)\biggr]\\
& - \lim_{h\to0}\mathbb{E}\left[\int_{z^h(t)}^{t}diag\left(V_{[\frac{t}{h}]}^{h}\right)\mu(V^h,s)\thinspace ds\right].
\end{align*}
For the last term above, since $\mu$ is bounded and $V_{[\frac{t}{h}]}^{h}=V^h(t)$, we have
\begin{align*}
\lim_{h\to0}\mathbb{E}\left[\left|\int_{z^h(t)}^{t}diag\left(V_{[\frac{t}{h}]}^{h}\right)\mu(V^h,s)\thinspace ds\right|\right]
 & \leq K\lim_{h\to0}h\mathbb{E}\left[\left|V^h(t)\right|\right]=0.
\end{align*}
 Denote by $\mathcal{N}_n^{\mu,h}:=\mathcal{N}^{\mu}(V^h,nh)$. Using  the Euler recursive formula \eqref{eq:V_n+1^h} and Assumption~\ref{a:bs}, we have
\begin{align*}
 & \left|\mathbb{E}\left[\sum_{n=1}^{[\frac{t}{h}]}\left(\left(V_{n}^{h}-V_{n-1}^{h}\right)-\int_{\left(n-1\right)h}^{nh}diag\left(V_{n-1}^{h}\right)\mu(V^h,s)\thinspace ds\right)\right]\right|\\
 =&\left| \mathbb{E}\left[\sum_{n=1}^{[\frac{t}{h}]}\left(
\int_{\left(n-1\right)h}^{nh}diag\left(V_{n-1}^{h}\right)(\mu^h_{n-1}-\mu(V^h,s))\thinspace ds\right)\right]\right|\\
\leq  &  \sum_{n=1}^{[\frac{t}{h}]}h\mathbb{E}\left[ \left\vert V_{n-1}^{h}\right\vert \left(Kh^{1/2} + \sum_{k=\mathcal{N}_{n-1}^{\mu,h}}^{\mathcal{N}_n^{\mu,h}}\left|J_{k}^{\mu}(V^h)\right|\right)\right]\\
\leq & Kh^{3/2} \sum_{n=1}^{[\frac{t}{h}]} \mathbb{E} \left\vert V_{n-1}^{h}\right\vert
+  Kh\sum_{n=1}^{[\frac{t}{h}]}\mathbb{E}\left[\left\vert V_{n-1}^{h}\right\vert\ [\mathcal{N}_n^{\mu,h}-\mathcal{N}_{n-1}^{\mu,h}]\right]. \numberthis \label{eq:E_M_t}\\
 \leq & Kh^{3/2} \sum_{n=1}^{[\frac{t}{h}]} \mathbb{E}\left[\sup_{0\leq t\leq T}\left\vert V^h(t)\right\vert \right] +
Kh\mathbb{E}\left[\sup_{0\leq t\leq T}\left\vert V^h(t)\right\vert \sum_{n=1}^{[\frac{t}{h}]}[\mathcal{N}_n^{\mu,h}-\mathcal{N}_{n-1}^{\mu,h}]\right]\\
 \leq & Kh^{1/2}\mathbb{E}\left[\sup_{0\leq t\leq T}\left\vert V^h(t)\right\vert \right]
+ Kh\mathbb{E}\left[\sup_{0\leq t\leq T}\left\vert V^h(t)\right\vert \right]
\end{align*}
which  tends to 0 as $h\to 0$  from (\ref{eq:mod12}). Here we have used  the fact that $\mathcal{N}^{\mu}(V^h,\cdot)$ is uniformly bounded given Assumption~\ref{a:bs} in the last inequality.
Therefore, $\mathbb{E}\left[M\left(\overline{V}, t\right)\right]=0$.
Using the same procedure one can show $\mathbb{E}\left[M\left(\overline{V}, t\right)\vert\mathcal{F}_{s}\right]=M\left(\overline{V}, s\right)$
for any $0\leq s\leq t$, thus $M(\overline{V},\cdot)$ is a martingale.

Denote the $l$th component of the vector process $X$ by $X_l$, and let $\left\langle X_{l},X_{q}\right\rangle \left(t\right)$
be the cross-variation of two real processes $X_{l}$ and $X_{q}$
up to time $t$. For  $t\in [0,T]$ denote by
\[
{Q}_{lq}(V, t):= \left\langle V_{l},V_{q}\right\rangle \left(t\right)-\int_{0}^{t}\left(diag\left(V\left(s\right)\right)\sigma\left(V,s\right)\sigma\left(V,s\right)^{T}diag\left(V\left(s\right)\right)\right)_{lq}\thinspace ds.
\]
Again, due to the uniform topology of the convergence of $\widetilde{V}^h$ to $\hat{V}$,  boundedness of $\sigma^h$ and \eqref{eq:phi_uniform_conv},
and the uniformly integrability of $\{{Q}_{lq}(\widetilde V^h, t):h>0\}$ from \eqref{eq:mod12},
we have
\[
\E \big|{Q}_{lq}(\overline{V}, t)\big| = \E \big|{Q}_{lq}(\hat V, t)\big| =\lim_{h\to 0}\E \big|{Q}_{lq}(\widetilde V^h, t)\big|=\lim_{h\to 0}\E \big|{Q}_{lq}(V^h, t)\big|.
\]
Therefore
\begin{align*}
\E \big|{Q}_{lq}(\overline{V}, t)\big|=& \lim_{h\to0}\mathbb{E}\bigg|\sum_{n=1}^{[\frac{t}{h}]}\left(V_{l,n}^{h}-V_{l,n-1}^{h}\right)\left(V_{q,n}^{h}-V_{q,n-1}^{h}\right)\\
& \quad  - \int_{0}^{t}\left(diag\left(V^{h}\left(s\right)\right)\sigma(V^h,s)\sigma(V^h,s)^{T}diag\left(V^{h}\left(s\right)\right)\right)_{lq}\thinspace ds\bigg|\\
\leq & \lim_{h\to0}\E \bigg| \sum_{n=1}^{[\frac{t}{h}]}\left[(V_{l,n}^h-V_{l,n-1}^h)(V_{q,n}^h-V_{q,n-1}^h)-V_{l,n-1}^hV_{q,n-1}^h(\sigma_{n-1}^h(\sigma_{n-1}^h)^T)_{lq}h\right] \bigg| \\
&\quad +\lim_{h\to0} \E \bigg| \sum_{n=1}^{[\frac{t}{h}]} V_{l,n-1}^hV_{q,n-1}^h \int_{(n-1)h}^{nh}(\sigma(V^h,s)\sigma(V^h,s)^T-\sigma_{n-1}^h(\sigma_{n-1}^h)^T)_{lq}ds\bigg|\\
 & \quad+\lim_{h\to0}\mathbb{E}\biggl|\int_{z^h(t)}^{t}\left(diag\left(V_{[\frac{t}{h}]}^{h}\right)\sigma(V^h,s)\sigma(V^h,s)^{T}diag\left(V_{[\frac{t}{h}]}^{h}\right)\right)_{lq}\thinspace ds\biggr|.\numberthis \label{eq:two_limits}
\end{align*}

The second and third limit terms  in (\ref{eq:two_limits}) are bounded by
$$ \lim_{h\to0}
\left(\sum_{n=1}^{[\frac{t}{h}]}K h^{\frac{3}{2}}\E\left[ V_{l,n}^hV_{q,n}^h\right] +Kh \sum_{n=1}^{[\frac{t}{h}]} \E\left[V_{l,n-1}^hV_{q,n-1}^h[\mathcal{N}^{\sigma,h}_n-\mathcal{N}^{\sigma,h}_{n-1}]\right]\right)
+ \lim_{h\to0}\mathbb{E}\left|V_{l,[\frac{t}{h}]}^{h}V_{q,[\frac{t}{h}]}^{h}\right|
$$
which is zero from a
similar argument for showing \eqref{eq:E_M_t} goes to zero as $h\to 0$ and  \eqref{eq:mod12}.
For the first limit term in \eqref{eq:two_limits}, using Assumption~\ref{a:bs} and the Euler recursive formula \eqref{eq:V_n+1^h}, we have
\begin{align*}
 & \E \bigg| \sum_{n=1}^{[\frac{t}{h}]}\left[(V_{l,n}^h-V_{l,n-1}^h)(V_{q,n}^h-V_{q,n-1}^h)-V_{l,n-1}^hV_{q,n-1}^h(\sigma_{n-1}^h(\sigma_{n-1}^h)^T)_{lq}h\right] \bigg| \\
\leq &  \E \bigg|\sum_{n=1}^{[\frac{t}{h}]} V_{l,n-1}^hV_{q,n-1}^h\mu_{l,n-1}^h\mu_{q,n-1}^h h^2 \bigg| \\
&\quad + \E \bigg|\sum_{n=1}^{[\frac{t}{h}]}\bigg\{h^{\frac{3}{2}} V_{l,n-1}^hV_{q,n-1}^h\big[\mu_{l,n-1}^h\sum_{i=1}^{k+1}(\sigma_{n-1}^h)_{qi}(Z_n)_i+\mu_{q,n-1}^h\sum_{i=1}^{k+1}(\sigma_{n-1}^h)_{li}(Z_n)_i\big]\bigg\}\bigg|\\
&\quad + \E \bigg| \sum_{n=1}^{[\frac{t}{h}]} hV_{l,n-1}^h V_{q,n-1}^h \big\{\sum_{i=1}^{k+1}(\sigma_{n-1}^h)_{li}(\sigma_{n-1}^h)_{qi}[(Z_n)_i^2-1]+\sum_{1\leq i\neq j\leq k+1}(\sigma_{n-1}^h)_{li}(\sigma_{n-1}^h)_{qj}[(Z_n)_i(Z_n)_j] \big\} \bigg|\\
\leq &  Kh\E\bigg[\sup_{0\leq s\leq T}|V^h(s)|^2\bigg]+\E\bigg[K\sup_{0\leq s\leq T}|V^h(s)|^2h^{\frac{3}{2}} \bigg|\sum_{n=1}^{[\frac{t}{h}]}\sum_{i=1}^{k+1}(Z_n)_i\bigg|\bigg]\\
&\quad +\E\bigg[K\sup_{0\leq s\leq T}|V^h(s)|^2h \bigg|\sum_{n=1}^{[\frac{t}{h}]}\sum_{i=1}^{k+1}[(Z_n)_i^2-1]\bigg|\bigg]+\E\bigg[K\sup_{0\leq s\leq T}|V^h(s)|^2h \bigg|\sum_{n=1}^{[\frac{t}{h}]}\sum_{1\leq i\neq j\leq k+1}[(Z_n)_i(Z_n)_j]\bigg|\bigg]\label{eq:E_M^2} \numberthis
\end{align*}
The first term of \eqref{eq:E_M^2}  clearly converges to zero. We only show the second term converges to zero. The convergence of the other two terms to zero can be shown similarly. Thanks to \cite[Theorem 2.5.7]{Durrett2010} $\sum_{n=1}^{[\frac{t}{h}]}\sum_{i=1}^{k+1}(Z_n)_i$ is at the order of $(\frac{1}{h})^{1/2}(\log\frac{1}{h})^{1/2+\epsilon}, \epsilon>0$. Hence
\[
\lim_{h\to 0} h^{\frac{3}{2}} \bigg|\sum_{n=1}^{[\frac{t}{h}]}\sum_{i=1}^{k+1}(Z_n)_i\bigg| = 0,\quad \mbox{a.s.}
\]
Therefore once we can justify the exchange of limit and expectation, then the convergence of the expectation to zero is shown. To this end, using Lemma~\ref{le:Ros_ineq} and Young's inequality, we show uniform integrability of $\{\sup_{0\leq s\leq T}|V^h(s)|^2h^{\frac{3}{2}} \bigg|\sum_{n=1}^{[\frac{t}{h}]}\sum_{i=1}^{k+1}(Z_n)_i\bigg|,h>0\}$ as follows
\begin{align*}
\E\bigg[K\sup_{0\leq s\leq T}|V^h(s)|^2h^{\frac{3}{2}} \bigg|\sum_{n=1}^{[\frac{t}{h}]}\sum_{i=1}^{k+1}(Z_n)_i\bigg|\bigg]^2\leq & \E\bigg[K\sup_{0\leq s\leq T}|V^h(s)|^8\bigg]+K\E[h^6|\sum_{n=1}^{[\frac{t}{h}]}\sum_{i=1}^{k+1}(Z_n)_i|^4]\\
\leq & K+ h^6 \max(\sum_{n=1}^{[\frac{t}{h}]}\sum_{i=1}^{k+1}\E|(Z_n)_i|^4,(\sum_{n=1}^{[\frac{t}{h}]}\sum_{i=1}^{k+1}\E(Z_n)_i^2)^2) \\
\leq &K+ K\max(h^5,h^4),
\end{align*}
which is finite for all small $h$, here we have used Lemma~\ref{le:Ros_ineq} in the second last inequality above. Hence we show \eqref{eq:E_M^2} converges to zero as $h\to 0$. Therefore
\[
\left\langle\overline{V}_{l},\overline{V}_{q}\right\rangle \left(t\right) = \int_{0}^{t}\left(diag\left(\overline{V}\left(s\right)\right)\sigma\left(\overline{V},s\right)\sigma\left(\overline{V},s\right)^{T}diag\left(\overline{V}\left(s\right)\right)\right)_{lq}\thinspace ds,\quad \mbox{a.s.}
\]
Thanks to \cite[Theorem 7.1]{Ikeda1989}, there exists a $k+1$ dimensional Brownian motion $B$
such that $M\left(\overline{V}, t\right)=\int_{0}^{t}diag\left(\overline{V}\left(s\right)\right)\sigma\left(\overline{V},s\right)\thinspace dB\left(s\right)$. Therefore, $\overline{V}$ is the weak solution of \eqref{eq:dVt}.\hfill$\square$

\subsection{Properties of First Passage Times}
\begin{lemma}
\label{Le:S_map_cont}
For any $n\in\mathbb{N}$ and $j\in\left\{ 1,2,\dots,n\right\} $,
the mapping $x\protect\mapsto\mbox{\ensuremath{\mathcal{S}}}^{n}\left(x,j\right)$
from $\mathbb{R}^n$ to $\mathbb{R}$ is continuous.
\end{lemma}

\noindent
\textit{Proof:} The case of $n=1$ is trivial.
For $n\geq 2$ and $x=\left(x_{1},\dots,x_{n}\right)^{T} \in\mathbb{R}^{n}$,
functions
$$\mathcal{S}^{n}\left(x,1\right)=\min\left(x_{1},\ldots,x_{n}\right)
\mbox{ and }\mathcal{S}^{n}\left(x,n\right)=\max\left(x_{1},\ldots, x_{n}\right)
$$ are obviously  continuous.
For $j\in\left\{ 2,\dots,n-1\right\} $ (a nonempty set only if $n\geq 3$),
we may decompose $x$ into  $\hat{x}=\left(x_{1},\dots,x_{n-1}\right)^{T}$
and $x_{n}$, and  express $\mathcal{S}^{n}\left(x,j\right)$ as
\[
\mathcal{S}^{n}\left(x,j\right)=\min\left\{ \max\left(\mathcal{S}^{n-1}\left(\hat{x},j-1\right),x_{n}\right),\mathcal{S}^{n-1}\left(\hat{x},j\right)\right\} .
\]
Given $\mathcal{S}^{n-1}\left(\cdot,j\right)$ is continuous, and  $\mathcal{S}^{n}\left(\cdot,1\right)$
and $\mathcal{S}^{n}\left(\cdot,n\right)$ are continuous, we have $\mathcal{S}^{n}\left(\cdot,j\right)$
is continuous for any $j\in\left\{ 1,2,\dots,n\right\} $. \hfill$\square$

\begin{lemma}
\label{Le:zero_meas_1D}
Given Assumption \ref{a:bs} and \ref{a:nondegenerate} hold, then we have
\[
\mathbb{P}\left(\tau_{0}=t_{j}\right)=0\text{ and }\mathbb{P}\left(\tau_{\left(i\right)}=t_{j}\right)=0,
\]
for any $i=1,2,\dots,k$ and $j=1,2,\dots,m$.
\end{lemma}

\noindent
\textit{Proof:} Note that firm values $V_{i}$, $i=0,1,\dots,k$,
satisfy \eqref{eq:dVt}. For any $j=1,2,\dots,m$, we show $\mathbb{P}\left(\tau_{0}=t_{j}\right)=0$
as follows
\begin{align*}
\mathbb{P}\left(\tau_{0}=t_{j}\right) & \leq\mathbb{P}\left(V_{0}\left(t_{j}\right)=L_{0}\left(t_{j}\right)\right)\\
 & =\mathbb{P}\left(V_{0}\left(0\right)\exp\left(\int_{0}^{t_{j}}\left(\mu_{0}\left(s\right)-\frac{b_{0}^{2}\left(s\right)}{2}\right)\thinspace ds+\int_{0}^{t_{j}}b_{0}\left(s\right)\thinspace dB_{0}\left(s\right)\right)=K_{0}\exp\left(\gamma_{0}t_{j}\right)\right)\\
 & =\mathbb{P}\left(\int_{0}^{t_{j}}\left(\mu_{0}\left(s\right)-\frac{b_{0}^{2}\left(s\right)}{2}\right)\thinspace ds+\int_{0}^{t_{j}}b_{0}\left(s\right)\thinspace dB_{0}\left(s\right)=\ln\frac{K_{0}}{V_{0}\left(0\right)}+\gamma_{0}t_{j}\right),\numberthis  \label{eq:P_int_0^t_j}
\end{align*}
where $b_{0}^{2}(s)=(1,0,\dots,0)(\sigma(s)\sigma(s)^{T})(1,0,\dots,0)^{T}$.
Due to the uniform nondegeneracy Assumption~\ref{a:nondegenerate} and boundedness of $\sigma$, there exist some positive constants $\lambda$ and $\Lambda$ such that
\[
0<\lambda<b_0\left(t\right)<\Lambda\quad\text{for all \ensuremath{0\leq t\leq T},}\thinspace\thinspace\text{a.s.}.
\]
Thanks to the property discussed in \cite[Appendix A.5]{Leung2013},
we know the term in \eqref{eq:P_int_0^t_j} is zero. Similarly, $\mathbb{P}\left(\tau_{i}=t_{j}\right)=0$ for
$i=1,2,\dots,k$. Finally,
$\mathbb{P}\left(\tau_{\left(i\right)}
=t_{j}\right)\leq\sum_{i=1}^{k}\mathbb{P}\left(\tau_{i}=t_{j}\right)=0$.
\hfill$\square$

\begin{remark}
In the proof of Lemma \ref{Le:zero_meas_1D}, we need to show a one-dimensional continuous semi martingale never hits a given point within finite time almost surely, more specifically see \eqref{eq:P_int_0^t_j}. One may
wonder if uniform nondegeneracy of $\sigma$ can be relaxed to positive definite to show this result. The answer is negative. The counter example for one dimensional case is provided in \cite{MathOverflow}.
\end{remark}

\begin{lemma}
\label{Le:zero_meas_2D}
Suppose Assumption~\ref{a:nondegenerate} is true. Given Assumption \ref{a:zero_corr} or \ref{a:ps_const} holds, then we have
\[
\mathbb{P}\left(\tau_{0}=\tau_{\left(i\right)}\wedge T\right)=0,
\]
for any $i=1,2,\dots,k$.
\end{lemma}

\noindent
\textit{Proof:} Since $\left\{ \tau_{0}=\tau_{\left(i\right)}\wedge T\right\} \subseteq{\displaystyle \cup_{i=1}^{k}}\left\{ \tau_{0}=\tau_{i}\wedge T\right\} $,
we have
\[
\mathbb{P}\left(\tau_{0}=\tau_{\left(i\right)}\wedge T\right)\leq\sum_{i=1}^{k}\mathbb{P}\left(\tau_{0}=\tau_{i}\wedge T\right).
\]
Observe that for any $i=1,2,\dots,k$,
\begin{align*}
\mathbb{P}\left(\tau_{0}=\tau_{i}\wedge T\right) & \leq\mathbb{P}\left(\tau_{0}=\tau_{i}\text{ and }\tau_{i}\leq T\right)+\mathbb{P}\left(\tau_{0}=T\text{ and }\tau_{i}>T\right)\\
 & \leq\mathbb{P}\left(\text{2-d random process \ensuremath{\left\{ \left(V_{0}\left(t\right),V_{i}\left(t\right)\right)\right\} _{t}}\thinspace\thinspace hits the curve \ensuremath{\left(L_{0}\left(t\right),L_{i}\left(t\right)\right),}\ensuremath{0\leq t\le T}}\right)\\
 & =\mathbb{P}\left(\text{process \ensuremath{\left\{ \left(V_{0}\left(t\right)-L_{0}\left(t\right),V_{i}\left(t\right)-L_{i}\left(t\right)\right)\right\} _{t}}\thinspace\thinspace hits \ensuremath{\left(0,0\right)}\thinspace\thinspace at some time \ensuremath{t}\thinspace\thinspace in \ensuremath{\left[0,T\right]}}\right). \numberthis \label{eq:P_V0-L0}
\end{align*}
In the above, we use the fact $\mathbb{P}\left(\tau_{0}=T\text{ and }\tau_{i}>T\right)  = 0$.  Similar to the derivation of \eqref{eq:P_int_0^t_j}, the event $\left\{ V_{i}\left(t\right)-L_{i}\left(t\right)=0\right\} $
is equivalent to
\[
\left\{ \int_{0}^{t}b_{i}\left(s\right)\thinspace dB_{i}\left(s\right)-\ln\frac{K_{i}}{V_{i}\left(0\right)}-\gamma_{i}t+\int_{0}^{t}\left(\mu_{i}\left(s\right)-\frac{b_{i}^{2}\left(s\right)}{2}\right)\thinspace ds=0\right\} .
\]
Define the processes $\widetilde{V}_{0}$ and $\widetilde{V}_{i}$
to be:
\begin{eqnarray*}
\widetilde{V}_{j}\left(t\right) & := & \int_{0}^{t}b_{j}\left(s\right)\thinspace dB_{j}\left(s\right)-\ln\frac{K_{j}}{V_{j}\left(0\right)}-\gamma_{j}t+\int_{0}^{t}\left(\mu_{j}\left(s\right)-\frac{b_{j}^{2}\left(s\right)}{2}\right)\thinspace ds,\quad j=0,i.
\end{eqnarray*}
The differential form of the SDE for $\widetilde{V}=\left(\widetilde{V}_{0},\widetilde{V}_{i}\right)^{T}$
is as follows
\begin{align*}
d\widetilde{V}(t) & =\hat\mu_i(t) \thinspace dt+\hat{b}_{i}\left(t\right)
\hat{\rho}_{i}\left(t\right) \thinspace dB'(t),
\end{align*}
where
\begin{eqnarray*}
\hat\mu_i(t) =\left(\begin{array}{c}
\mu_{0}\left(t\right)-\frac{b_{0}^{2}\left(t\right)}{2}-\gamma_{0}\\
\mu_{i}\left(t\right)-\frac{b_{i}^{2}\left(t\right)}{2}-\gamma_{i}
\end{array}\right),\,
\hat{b}_{i}\left(t\right)  =  \begin{pmatrix}b_{0}\left(t\right) & 0\\
0 & b_{i}\left(t\right)
\end{pmatrix},\,
\hat{\rho}_{i}\left(t\right)  =  \begin{pmatrix}1 & 0\\
\rho_{i}\left(t\right) & \sqrt{1-\rho_{i}^{2}\left(t\right)}
\end{pmatrix},
\end{eqnarray*}
and $B^{'}=\left(B_{0}^{'},B_{i}^{'}\right)^{T}$
is a standard 2-d Brownian motion, $\rho_{i}$ is an adapted process satisfying
\[
\hat{b}_{i}\left(t\right)
\hat{\rho}_{i}\left(t\right)\hat{\rho}_{i}\left(t\right)^T\hat{b}_{i}\left(t\right)=\left(\begin{array}{c}
\sigma_{i}\left(t\right)\\
\sigma_{j}\left(t\right)
\end{array}\right)\left(\begin{array}{c}
\sigma_{i}\left(t\right)\\
\sigma_{j}\left(t\right)
\end{array}\right)^{T},
\]
and $\sigma_{i}\left(t\right)$ and $\sigma_{j}\left(t\right)$
are $i$th and $j$th row vector of $\sigma\left(t\right)$.  Rewriting \eqref{eq:P_V0-L0} in terms of the
process $\widetilde{V}$, our target is to show that if Assumption \ref{a:zero_corr} or \ref{a:ps_const} is true, then
\begin{equation}
\label{eq:PtildeV_zero}
\mathbb{P}\left(\widetilde{V}\left(t\right)=\left(0,0\right)\quad\text{for some \ensuremath{0<t\leq T}}\right)=0.
\end{equation}
Using Girsanov Theorem, we may assume the drift term is zero:
\begin{eqnarray*}
d\widetilde{V}\left(t\right) & = & \hat{b}_{i}\left(t\right)\hat{\rho}_{i}\left(t\right)\thinspace dB'\left(t\right):=\hat{\sigma}\left(t\right)\thinspace dB'\left(t\right).
\end{eqnarray*}
We first prove \eqref{eq:PtildeV_zero} is true under Assumption~\ref{a:zero_corr}. We have  $\hat{\rho}_{i}\left(t\right)$
is an identity matrix for all $t\geq0$ almost surely.
Since we are investigating the first-passage time before $T$, without
loss of generality, we assume
\[
b_{0}\left(t\right)\equiv1\text{ and }b_{i}\left(t\right)\equiv1,\quad\forall t>T,\text{ a.s.}
\]
Then we have
\[
\text{\ensuremath{\left\langle \widetilde{V}_{0},\widetilde{V}_{0}\right\rangle(\infty)}=\thinspace\thinspace\ensuremath{\infty}, \ensuremath{\left\langle \widetilde{V}_{i},\widetilde{V}_{i}\right\rangle(\infty)}=\thinspace\thinspace\ensuremath{\infty}}\thinspace\thinspace\text{and }\left\langle \widetilde{V}_{0},\widetilde{V}_{i}\right\rangle (t)=\thinspace\thinspace0,\thinspace\thinspace\forall t>0,\text{ a.s.}
\]
Define $\hat{T}_0(t):=\inf\{s:\langle\widetilde{V}_0,\widetilde{V}_0\rangle(s)>t\}$ and $\hat{T}_i(t):=\inf\{s:\langle\widetilde{V_i},\widetilde{V_i}\rangle(s)>t\}$. By the time change results for multidimensional continuous local martingales in \cite[Theorem 5.1.9]{Revuz1999}, we have
\[
\hat{B}\left(s\right):=\left(\widetilde{V}_{0}\left(\hat{T}_{0}\left(s\right)\right),\widetilde{V}_{i}\left(\hat{T}_{i}\left(s\right)\right)\right)^{T},\quad s\geq0,
\]
is a standard 2-d Brownian motion under $\mathbb{P}$ with initial position
$\left(\widetilde{V}_{0}\left(0\right),\widetilde{V}_{i}\left(0\right)\right)$
. Since $\langle\widetilde{V_0},\widetilde{V_0}\rangle(\cdot)$ and $\langle\widetilde{V_i},\widetilde{V_i}\rangle(\cdot)$ are both continuous and strictly increasing, the inverse maps $\hat{T}^{-1}_0$ and $\hat{T}^{-1}_i$ exist and are both strictly increasing. It suffices to show
\[
\mathbb{\mathbb{P}}\left(\hat{B}\left(t\right)\text{ hits origin\thinspace\thinspace at some \ensuremath{t}\thinspace\thinspace in \ensuremath{\left(0,\max\left(\hat{T}^{-1}_{0}\left(T\right),\hat{T}^{-1}_{i}\left(T\right)\right)\right]}}\right)=0.
\]
This is true due to $\max\left(\hat{T}^{-1}_{0}\left(T\right),\hat{T}^{-1}_{i}\left(T\right)\right)<\infty,a.s.$,  and nonattainability of the origin by the 2-d Brownian
path shown in \cite[Proposition 3.3.22]{Karatzas1991}.

We next prove \eqref{eq:PtildeV_zero} is true under Assumption \ref{a:ps_const}. We have $\hat{\sigma}$ is a piecewise constant process almost surely. In other words, for  strictly increasing
stopping time sequence $\left\{ \theta_{0},\theta_{1},\dots\right\} $
such that $\theta_0 = 0$ and $\lim_{n\to\infty}\theta_n = 
T+1$
, the process $\hat{\sigma}$
is in the form of
\[
\hat{\sigma}\left(t\right)=\sum_{i=1}^{\infty}\hat{\sigma}_{i}\mathbf{1}\left(\theta_{i-1}
\le t < \theta_{i}\right),\quad\text{a.s.,}
\]
where $\hat{\sigma}_{i}$ is a nonsingular $2\times2$
dimensional matrix, measurable with respect to $\mathcal{F}_{\theta_{i-1}}$.
Let $p$ be any given point in $\mathbb{R}^{2}$. Partitioning the
event in \eqref{eq:PtildeV_zero}, we have
\begin{align*}
 & \mathbb{P}\left(\widetilde{V}(t)=p\quad\text{for some \ensuremath{0<t\leq T}}\right)\\
\leq \thinspace\thinspace & \sum_{i=1}^{\infty}\mathbb{P}\left(\widetilde{V}(t)=p\quad
\text{for some \ensuremath{\theta_{i-1}\le t< \theta_{i}}}\thinspace\thinspace\text{\ensuremath{< T + 1}}\right)\\
= \thinspace\thinspace & \sum_{i = 1}^{\infty} \int_{\mathbb{R}^{2}}\mathbb{P}\left(\widetilde{V}(t)=p\quad
\text{for some \ensuremath{\theta_{i-1}\le t< \theta_{i}}}\thinspace\thinspace\text{\ensuremath{< T + 1}}
\biggr|\widetilde{V}(\theta_{i-1})=x\right)\mathbb{P}\left(\widetilde{V}(\theta_{i-1})\in dx\right).
\end{align*}
But the integrand in the above integral is seen to be zero from the following
derivation.
\begin{align*}
 & \mathbb{P}\left(\widetilde{V}(t)=p\quad
 \textrm{ for some } \theta_{i-1} \le t < \theta_{i} < T+1
  \biggr|\widetilde{V}(\theta_{i-1})=x\right)\\
=\thinspace\thinspace & \mathbb{P}\left(\int_{\theta_{i-1}}^{t}\hat{\sigma}_{i}\thinspace dW(s)=p- \widetilde{V}(\theta_{i-1})\quad
\textrm{ for some } \theta_{i-1} \le t < \theta_{i} < T+1
\biggr|\widetilde{V}(\theta_{i-1})=x\right)\\
\leq\thinspace\thinspace & \sup_{C}\mathbb{P}\left(\int_{\theta_{i-1}}^{t}C\thinspace dW(s)=p-x\quad
\text{for some \ensuremath{\theta_{i-1}\le t < \theta_{i} < T+1}
\ensuremath{\biggr|\widetilde{V}(\theta_{i-1})=x}}\right)\\
=\thinspace\thinspace & \sup_{C}\mathbb{P}\left(W(t)-W(\theta_{i-1})=C^{-1}\left(p-x\right)\quad\mbox{for some \ensuremath{\theta_{i-1} \le t< \theta_{i} < T+1}}\biggr|\widetilde{V}(\theta_{i-1})=x\right)\\
\leq\thinspace\thinspace & \sup_{C}\mathbb{P}\left(\hat W(t)=C^{-1}\left(p-x\right)\quad\text{for some \ensuremath{0<t\leq T+1}}\right)\\
=\thinspace\thinspace & 0.
\end{align*}
Here the supremum is taken among any constant 2$\times$2 nonsingular
matrix, since $\hat{\sigma}_{i}$ is measurable with respect to $\mathcal{F}_{\theta_{i-1}}$
and constant from $\theta_{i-1}$ to $\theta_{i}$. Also, $\hat W$ denotes a 2-dimensional
Brownian motion starting from zero. The last inequality is using
Markovian property of Brownian motion and $\theta_{i}-\theta_{i-1}$ is no greater
than $T+1$ almost surely. Hence for each $i$ the integral is zero, so is the countable summation. \hfill$\square$

Assumption \ref{a:zero_corr} or \ref{a:ps_const}
is used to prove \eqref{eq:PtildeV_zero}. Although it is intuitive to think the two dimensional
continuous nondegenerate local martingale (without Assumption~\ref{a:zero_corr} or \ref{a:ps_const}) should not hit a given
point within finite time, it seems difficult to prove this rigorously. Since this unsolved question is interesting on its own, we describe it in detail below.

\begin{remark}\label{r:oq}
(an open question)  Let $W$ be a two dimensional standard Brownian motion with respect
to a filtered probability space
$(\Omega, \mathcal F, \mathbb P, \mathbb F := \{\mathcal F_{t}\}_{t\ge 0})$. Define a continuous local martingale $Y$ by
$$
Y(t) = Y(0)+\int_{0}^{t} \sigma(s) d W(s),\quad t\geq 0
$$
with $Y(0)\neq 0$. Assume that $\sigma$ is an adapted $2\times 2$ matrix process satisfying
$$ \lambda \|\xi\|^{2} \le \xi^T \sigma(t) \sigma(t)^T \xi \le \Lambda  \|\xi\|^{2},
\hbox{ for all } (\xi, t) \in \mathbb R^{2}  \times (0,T) \hbox{ with some } \lambda, \Lambda >0, \hbox{ a.s.}$$
The question we have is that whether the following equality
$$\mathbb P\{Y(t) = (0,0), \ \hbox{ for some } t\in [0,T]\}=0$$
 is true? We leave this to our future work.
\end{remark}

\subsection{Completion of the Proof}
Now we are ready to complete the proof of Theorem~\ref{t:main}.
Let $C_{F_1}$ and $C_{F_2}$ be the sets of the continuities of functions $F_1$ and $F_2$ respectively. Due to \cite[Lemma 4]{Song2013}, with fixed boundary $L_{i}$,
the mapping $\pi\left(\cdot,L_{i}\right)$ is continuous at each $x\in C_{1}^{i}\cup C_{2}^{i}$. Notice that $F_1$ and $F_2$ are composition of indicator functions, $\mathcal{S}^k(\cdot,j)$ and $\pi(\cdot,L_i)$, and $\mathcal{S}^k(\cdot,j)$ is continuous by Lemma~\ref{Le:S_map_cont}. Hence we have
\begin{align*}
\widetilde{C}_{F_1}:=&\{x\in\C^{k+1}[0,T]: x_i\in C^i_1\cup C_2^i,i=0,\dots,k, \mbox{ and }\mathcal{S}^k(\{\pi(x_n,Ln\}_{n=1}^k,i)\neq T,\\
&\mbox{ and } \pi(x_0,L_0)\neq \mathcal{S}^k(\{\pi(x_n,L_n)\}_{n=1}^k,i)\wedge T\}\subset C_{F_1}
\end{align*}
and
\begin{align*}
\widetilde{C}_{F_2}:=&\{x\in\C^{k+1}[0,T]: x_i\in C^i_1\cup C_2^i,i=0,\dots,k, \mbox{ and}\\
&\mathcal{S}^k(\{\pi(x_n,Ln\}_{n=1}^k,i)\neq t_j,\mbox{ and } \pi(x_0,L_0)\neq t_j\}\subset C_{F_2}.
\end{align*}
To apply the mapping theorem (see \cite[Theorem 2.7]{Billingsley2009}) we need to show $\P(\omega:V(\omega)\in C_{F_1})=1$ and $\P(\omega:V(\omega)\in C_{F_2})=1$. It can be seen by
\begin{align*}
&\P(\omega:V(\omega)\in C_{F_1})\\
\geq&\P(\omega:V(\omega)\in \widetilde{C}_{F_1})\\
=&\P(V_i\in C_1^i \cup C_2^i, i=0,\dots,k, \mbox{ and }\mathcal{S}^k(\{\pi(V_n,Ln\}_{n=1}^k,i)\neq T,\\
&\mbox{ and } \pi(V_0,L_0)\neq \mathcal{S}^k(\{\pi(V_n,L_n)\}_{n=1}^k,i)\wedge T)\\
=&\P(V_i\in C_1^i \cup C_2^i, i=0,\dots,k, \mbox{ and }\tau_{(i)}\neq T,\mbox{ and } \tau_0\neq \tau_{(i)}\wedge T)\\
=&\P(\tau_{(i)}\neq T,\mbox{ and } \tau_0\neq \tau_{(i)}\wedge T) \mbox{   (Because $\P(V_i\in C_1^i \cup C_2^i, i=0,\dots,k)=1$}\\
& \quad \quad \quad \quad \quad \quad \quad \quad\quad \quad\quad \quad  \mbox{due to Assumption~\ref{a:nondegenerate} and \cite[Proposition A.1]{Song2012}})\\
=& 1. \quad \quad \mbox{(By Lemma~\ref{Le:zero_meas_1D} and Lemma~\ref{Le:zero_meas_2D})}
\end{align*}
Similarly, one can show $\P(\omega:V(\omega)\in C_{F_2})=1$.  With Assumptions~\ref{a:bs} and \ref{a:cont_phi}, Theorem~\ref{thm:V_wk_conv} says that $V^{h}\Rightarrow V$
as $h\to0$. Using the mapping theorem, we conclude that
\[
F_{i}\left(V^{h}\right)\Rightarrow F_{i}\left(V\right),\quad i=1,2,\quad\text{as \ensuremath{h\to0}}.
\]
Note that $F_{i}$, $i=1,2$, are bounded functions, hence $\{F_i(V^h):h>0\}$ are families of uniformly integrable random variables and
\[
\lim_{h\to0}\mathbb{E}\left[F_{i}\left(V^{h}\right)\right]=\mathbb{E}\left[F_{i}\left(V\right)\right],\quad i=1,2.
\]
Therefore, $\lim_{h\to0}
\hat{c}_i^h
=
\hat{c}_i$
\hfill$\square$

\begin{remark} (Extension to stochastic interest rate)
\label{re:stochastic_r}
Suppose the risk-free interest rate $r$ is a $\mathcal{F}_t$-adapted continuous process. Let $X$ be a $k+2$ dimensional $\mathcal{F}_t$-adapted continuous process which represents $V_i$, $i=0,1,\dots,k$ and $r$. Define $\mathcal{I}_1:\mathbb{D}^{k+2} \times 
\mathbb{N}\mapsto \mathbb{R}$ and 
$\mathcal{I}_2:\mathbb{D}^{k+2} 
\times [0,T] \mapsto \mathbb{R}$ as
\[
\mathcal{I}_1 (x,i):=\exp\left\{-\int_0^{S^k(\{\pi(x_n,L_n)\}_{n=1}^k,i)} x_{k+1}(u) du\right\},
\]
and
\[
\mathcal{I}_2 (x,t):=\exp\left\{-\int_0^{t} x_{k+1}(u) du\right\}.
\]
For $x_n\in \D$ and $x\in\C$ we have that $x_n$ converging to $x$ in Skorohod metric implies $x_n$ converging to $x$ in uniform metric which implies
$\int_0^t x_n(u)du$ converging to $\int_0^t x(u)du$ for $0\leq t\leq T$. This shows the mapping from $\D$ to $\mathbb{R}$, defined by
$x\to \int_0^t x(u)du$, is continuous at each $x\in \C$.
Then, the two mappings $\mathcal{I}_1(\cdot,i)$ and $\mathcal{I}_2(\cdot,t)$ are continuous at $\{x\in \C^{k+1}:x_i\in C_1^i \cup C_2^i,i=0,1,\dots,k\}$ under Skorohod topology. Since the discounting factors $exp\{-\int_0^{\tau_{(i)}} r_u du\}$ and $exp\{-\int_0^{t_j} r_u du\}$ can be rewritten as $\mathcal{I}_1(X,i)$ and $\mathcal{I}_2(X,t_j)$ respectively, Theorem \ref{t:main} still holds for the process $X$ under the same setting.
\end{remark}

\section{Conclusions}
We have derived the sufficient conditions for the convergence of the approximation of basket CDS with counterparty risk under a credit contagion model of multinames by generalizing the known weak limit theorems with discontinuous coefficents under non-Markovian setting. The method developed in this paper may be used to study other problems involving running maximal processes of correlated Brownian motions, the joint distribution of which is still unknown for dimension greater than two.

\bigskip\noindent
  {\bf Acknowledgement} The authors thank Thomas G. Kurtz for useful discussions on weak convergence of stochastic processes.
  The authors are very grateful to the anonymous reviewers and the AE whose instructive comments and suggestions have helped greatly to improve the paper of the previous version.


\end{document}